\documentclass[]{aa}  
\usepackage{graphicx, lscape,txfonts}
\def\lsim{\mathrel{\rlap{\lower4pt\hbox{\hskip1pt$\sim$}}
    \raise1pt\hbox{$<$}}}                
\begin{document}
   \title{A serendipitous survey for  variability amongst the massive  stellar
population of Westerlund 1\thanks{Based on observations collected at the European Southern Observatory,
Paranal and La Silla Observatories under programme IDs ESO 067.D-0211, 069.D-0039, 071.D-0151, 073.D-0327, 
075.D-0388, 081.D-0324 and 383.D-0633.}}

   \author{J.~S.~Clark\inst{1} \and B.~W.~Ritchie\inst{1,2} \and I.~Negueruela\inst{3}}
   \offprints{J.S.~Clark, \email{jsc@star.ucl.ac.uk}}

   \institute{
        Department of Physics and Astronomy, The Open University, Walton Hall, 
        Milton Keynes MK7 6AA, United Kingdom
   \and 
       IBM United Kingdom Laboratories, Hursley Park, Winchester, Hampshire
       S021 2JN, United Kingdom
   \and
        Departamento de F\'{\i}sica, Ingenier\'{\i}a de Sistemas y 
        Teor\'{\i}a de la Se\~{n}al, Universidad de Alicante,
        Apdo. 99, 03080 Alicante, Spain
   }

   \date{Accepted ??? Received ???}

   \abstract
{}
{Massive stars are known to demonstrate significant spectroscopic and  photometric variability over a wide range of 
   timescales. However the physical mechanisms driving this behaviour remain poorly understood.
   Westerlund 1 presents an ideal  laboratory for  studying  these processes in a rich,
coeval population of post-main sequence stars and we present a  pathfinding study aimed at 
    characterising their variability.}
   {To do this we utilised the large body of spectroscopic and photometric data that has accumulated for Wd1 during the
    past decade of intensive studies, supplemented with the sparser historical observations extending back to the early 
1960s.}
  {Despite the heterogeneous nature of this dataset, we were able to identify both spectroscopic and photometric variability 
    amongst every class of evolved massive star present within Wd~1. Spectroscopic variability attributable to both 
wind asphericity and photospheric pulsations was  present amongst both the hot and cool hypergiants and the former, also with
 the Wolf Rayets. Given the limitations imposed by the data, we
 were unable to determine the physical origin of the wind structure inferred for the OB 
supergiants, noting that it was present in both single pulsating and binary stars.  
In contrast we suspect that the inhomogineities in the winds of the Wolf Rayets are
 driven by binary interactions and, conversely, by pulsations in at least one of the cool hypergiants. 
Photospheric pulsations  were found for stars ranging from spectral types as early as O9 I
 through to the mid F Ia$^+$ yellow hypergiants - with a possible dependence on the luminosity class 
amongst the hot supergiants. The spectroscopically variable  red supergiants (M2-5 Ia) are also potential pulsators but 
require further observations to confirm this hypothesis. Given these findings it 
was therefore rather surprising that,  with the exception of W243, no  evidence of the characteristic
 excursions  of both luminous blue variables and 
 yellow hypergiants was found. Nevertheless, future determination of the 
amplitude  and periodicity of these pulsations as a function of temperature, luminosity and evolutionary state holds out 
the tantalising possibility of constraining the nature of the physical mechanisms driving the instabilities that constrain and 
define  stellar evolution in the upper reaches of the HR diagram. Relating to this,  the lack of secular evolution amongst the 
cool hypergiants and the presence of both high-luminosity yellow hypergiants  and red supergiants within Wd1 potentially place strong constraints on
 post-main sequence  evolutionary pathways, with the latter result apparently contradicting current theoretical predictions for
$>25$M$_{\odot}$ stars at solar metallicites.}
{}

  \titlerunning{Stellar variability in Wd1}
  \maketitle
%

\section{Introduction}

A large body of observational evidence exists that massive post-main sequence (MS) stars are subject to 
a variety of (pulsational) instabilities (e.g. Humphreys \& Davidson \cite{hd94}, Sterken \cite{sterken77}, Burki 
\cite{burki}, de Jager \cite{dj98}, Wood et al. \cite{wood}). Indeed, it has been 
supposed that such instabilities contribute to  the absence of cool (T$<10$kK) hypergiants above log(L/L$_{\odot}$)$\sim$5.8, giving rise 
to the  the Humphreys Davidson limit that spans the upper reaches of the HR diagram.

Characterisation of such instabilities is important for two reasons. Firstly, one might hope to use the pulsational properties to investigate 
the internal  structure of massive stars via asteroseismology. Secondly, there is increasing reason to suppose that MS mass loss
 is insufficient to permit the transition of high-mass stars  from unevolved O-type objects to H-depleted Wolf Rayet stars, and instead that the H-rich mantle is
 shed in short-lived impulsive events associated with the `zoo' of transitional objects. Indeed observational evidence of  this hypothesis is 
provided by the presence of fossil ejecta surrounding  luminous blue variables (LBVs: e.g. Clark et al. \cite{c03}), yellow 
hypergiants (YHGs; Castro-Carrizo et al. \cite{castro}) and red supergiants (RSGs; Schuster et al. \cite{schuster}) and  
the direct association of dramatic increases in  mass loss with episodes of 
instability in both hot and cool luminous stars ($\eta$ Car and  \object{$\rho$ Cas}, respectively; Smith \& Gehrz \cite{smith98}, Lobel et al. \cite{lobel}). Thus, determining the duration, duty cycle and mass loss rate of such events appears to be  necessary for the construction of a theory of massive 
stellar 
evolution.

These  instabilities are manifest in both  photometric and spectroscopic  variability, but other physical processes, such as
 large and small
scale wind inhomogineities, also lead to similar behaviour. Line driven winds are intrinsically unstable, leading to stochastic 
clumping (e.g. Owocki \cite{ow2000}),
 while hydrodynamical simulatons by Cranmer \& Owocki (\cite{cranmer}) demonstrate that large-scale photospheric perturbations 
- originating in non-radial pulsations
and/or magnetic fields  - can lead to large scale rotating wind structures. Moreover, spectrophotometric variability is an 
important observational signature of  wind collision zones in 
massive binaries (e.g. Lewis et al. \cite{lewis}, Stevens \& Howarth \cite{stevens}).
Of these phenomena, the degree of wind clumping present in massive stellar winds is  critical to  the observational 
determination of the  mass loss rate, while the characterisation of 
the properties of the massive binary population  is important for fields as diverse as star formation, the nature of SNe 
progenitors  and the formation channels of X-ray binaries.

Consequently, considerable observational effort has been expended in the identification and characterisation of variability 
in massive (binary) stars. Such  studies
 have typically centred on field stars, with the attendant difficulties of the determination of stellar luminosity and hence 
mass and age. However the development
of multiplexing spectrographs and the  identification of a large number of massive young Galactic clusters - of which 
Westerlund 1 is an examplar  (henceforth Wd1) - presents a unique 
opportunity to advance this field, offering  a large coeval population of massive stars of identical 
initial metallicity and well constrained masses.

First identified by Westerlund (\cite{westerlund61}), Wd1 was subject to sporadic optical and near-IR photometric studies in the 
following four decades (Sect. 2), with 
the high reddening towards it preventing a spectroscopic study until that of  Westerlund (\cite{westerlund87}; West87). However,
 despite the discovery of a remarkable population of  both early and late 
super/hypergiants, the cluster languished in relative obscurity until the turn of the century, when modern spectroscopic 
studies - motivated by the unexpected {\em radio} properties of Wd1 - first identified
a rich population of Wolf-Rayets (WRs; Clark \& Negueruela \cite{c02}) and subsequently revealed Wd1 to be the first example 
of a super star cluster  identified in the Galaxy (Clark et al. \cite{c05}; henceforth C05). 
In the following 
decade, Wd1 has been the subject of a large number of studies from X-ray to radio wavelengths aimed at determining both
 bulk cluster properties and those of the individual constituent stars.
The combined   dataset therefore offers the potential of investigating the  variability of cluster members over a significant 
time period ($\geq$40~yr for the brightest stars; Sect. 2) and we present
the results of such an analysis in this manuscript. In Sect. 2 we present and describe the reduction of the new data included 
in this work and discuss the utilisation of data available in previous 
studies. In Sect. 3-7 we analyse the available data as a function of evolutionary state and provide a discussion and summary 
in Sect. 8.


\section{Observations \& Data Reduction} \label{sec:obs_data}

\begin{figure}
\begin{center}
\includegraphics[width=7.55cm]{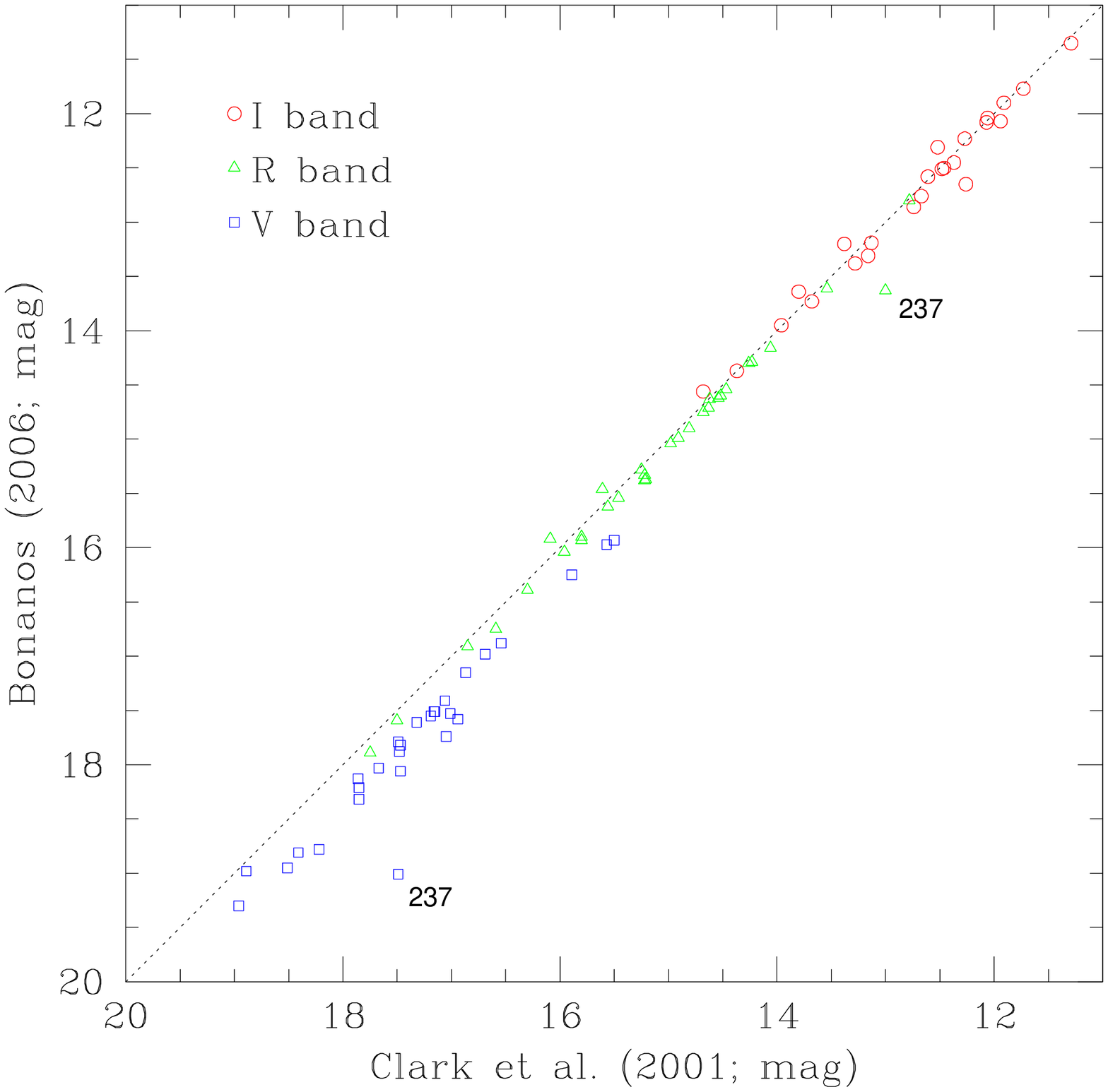}\\
\includegraphics[width=7.55cm]{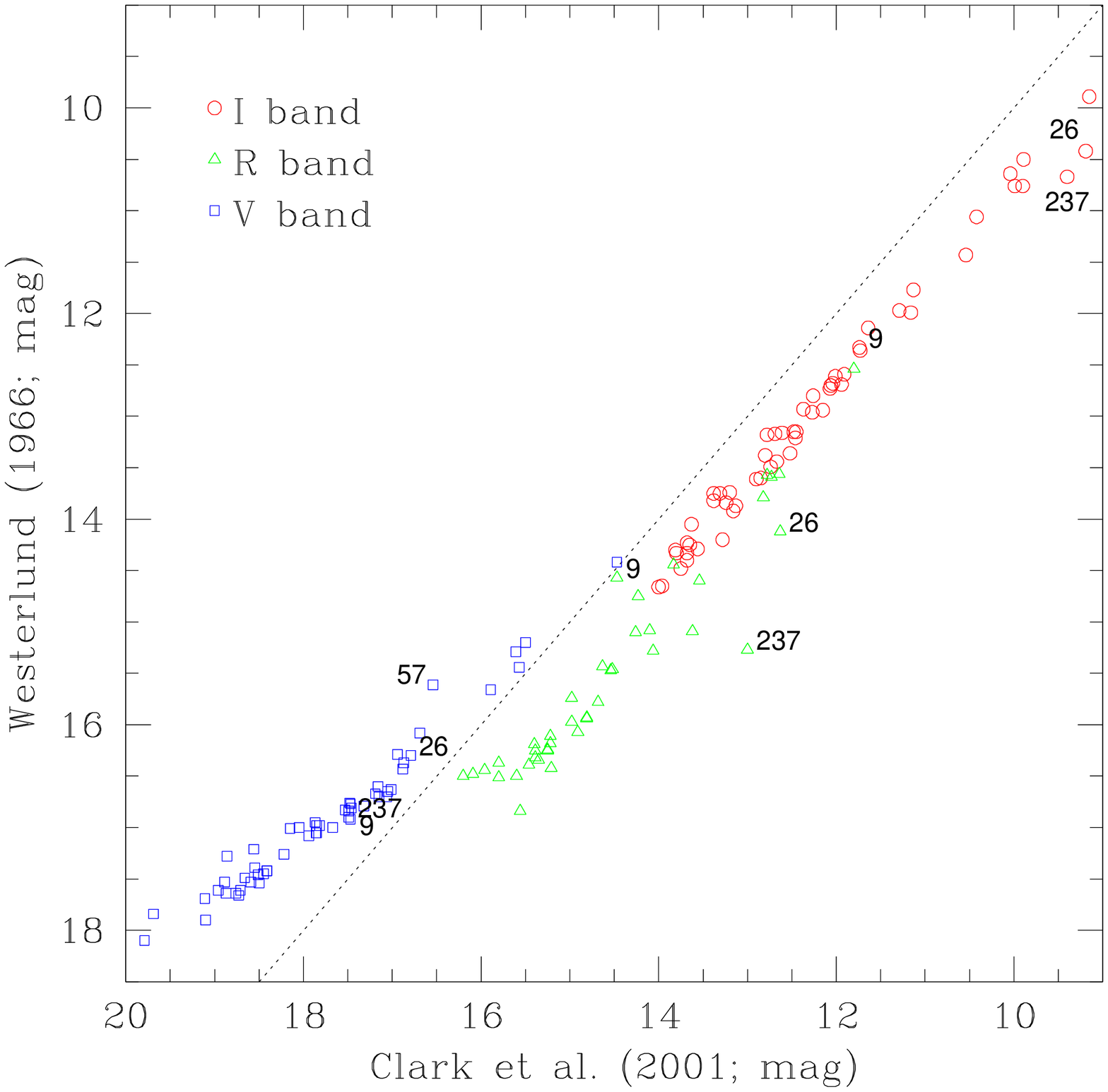}\\
\includegraphics[width=7.55cm]{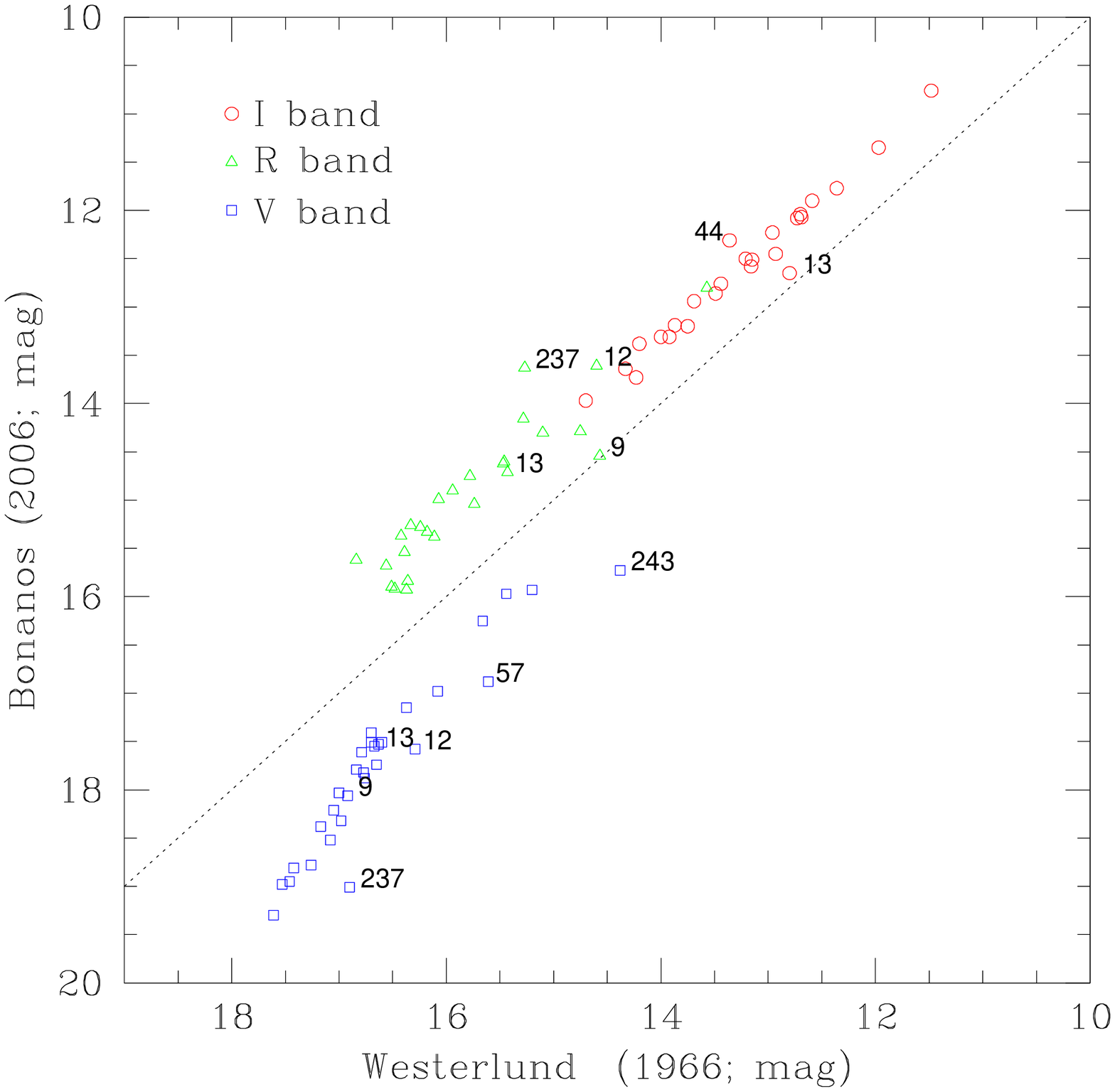}\\
\caption{Comparison of the photometric datasets presented by C05, Bonanos (\cite{bonanos})  and West87. Note  the zero 
point offsets present between all three datasets. In all three panels outliers are flagged, along with observations of such  stars in the other wavebands, where available. 
Also note that, for a given star,  photometry is not always available for all three datsets
(Sect. 2.1).}  
\end{center}
\end{figure} 

Between its discovery in  1961 and the era of intensive observations post 2001, 
Wd 1  was the subject of a number of photometric and spectroscopic studies ranging from optical 
to mid-IR wavelengths; these are summarised in Table 1. 
Unfortunately, in many cases observations collected in single works were not contemporaneous, complicating 
analysis; an important example is the photometry and spectroscopy presented by West87, which were obtained 
15 years apart. 
An additional difficulty  in comparing  multiepoch photometry is the different filter 
and detector (photographic plate, photometer and CCD) combinations employed by the various studies, for which 
 accurate spectral responses were unavailable. 
Combined with different extraction techniques (important due to the crowded nature of the cluster) and the lack
 of appropiate red standards, we found the number of observationally robust results that could be drawn from 
the historical  data were limited. 
Finally, the recent near-IR photometric observations of  Brandner et al. 
(\cite{brandner}) provide no extra information due to the saturation of the near-IR bright supergiants. 

\begin{table}
\begin{center}
\caption{Summary of published  observations utilised in this study}
\begin{tabular}{lll}
\hline
\hline
Year  & Dataset & Reference \\
\hline
1961 &  VI band phot. & Borgman et al. (\cite{borgman}) \\
1966 &  VRI band phot. & West87 \\
1969    & K band phot. & Borgman et al. (\cite{borgman}) \\
1973    & V+narrow band phot. & Lockwood (\cite{lockwood74}) \\
1973-4   & 2-20$\mu$m phot. & Koornneef (\cite{koor}) \\
1995    & VI band phot. & Piatti et al. (\cite{piatti}) \\
 2001   & BVRI band phot& C05 \\
2006    & BVRI band  phot. & Bonanos (\cite{bonanos}) \\
\hline
1981    & RI band spec. & West87 \\
2001    & RI band spec. & C05 \\
2002    & RI band spec. & C05 \\
2003    & RI band spec. & Negueruela \\
        &               & \& Clark (\cite{nc})\\
2005    & IJ \& HK spec. & Crowther et al. (\cite{crowther}) \\
2006   & K band spec. & Mengel \& \\
       &              & Tacconi-Garman (\cite{mengel})\\
\hline
\end{tabular}
\end{center}
\end{table}
 
\begin{table*}
\begin{center}
\caption{Comparison of previously published spectral classifications.}
\begin{tabular}{lccccccc}
\hline
\hline
W\# &     \multicolumn{7}{c}{Spectral Classification} \\
    & 1966  & 1973 & 1973-4? & 1981 & 1995 &2002  & 2006\\
\hline
 4  &{\em OB}& {\em OB}&-       &G0 Ia & {\em OB} &F2 Ia & $<$G5 Ia \\
 6  &{\em OB}& -       &-       &B1 V  &   -      & B0.5 Iab & -\\
 7  &{\em OB}& {\em OB}&-       &A0 Ia & {\em OB} & B5 Ia & - \\
 8a &{\em OB}& -       &-       &G0 Ia & {\em OB} & F5 Ia & $<$G5 Ia \\
 9  &{\em OB}& ?       & ?      &Be    & {\em OB} & sgB[e] & $<$G5 Ia \\
10  &{\em OB}& -       &-       &B0 III&   -      & B0.5I+OB & -\\
11  &{\em OB}& -       &-       &B1 II &   -      & B2 Ia & -\\
12a  &{\em OB}&{\em OB} &-       &A2 Ia & {\em OB} & A5 Ia & $<$G5 Ia \\
16a  &{\em OB}&{\em OB} &-       &A2 Ia & {\em OB} & A2 Ia & -\\
20  &{\em M} &{\em M}  &{\em M} &M6 Ia &  -       & $<$M6 I & M5 Ia\\
26  &{\em M} &{\em M}  &{\em M} &M2 Ia & {\em M}  & $<$M6 I & M5-6Ia\\
28  &{\em OB}& -       & -      &B0 II &  -       &B2 Ia & - \\
31  &{\em OB}& -       & -      &Be    &  -       &B0I+OB & - \\
32  &{\em OB}&{\em OB} & -      &G5 Ia & {\em OB} &F5 Ia & $<$G5 Ia \\
33  &{\em OB}& -       & -      &B8 Iab& {\em OB} &B5 Ia & -\\
70  &{\em OB}& -       & -      &B8 Iab&  -       & B3 Ia & - \\
71  &{\em OB}& -       & -      &B8 Iab&  -       & B2.5 Ia & -\\
75  &{\em M}&{\em  M} &{\em M} &-     &  -       & -  & M4 Ia\\
78  &{\em OB}& -       &-       &B1 Ia &  -       & B1 Ia & -\\
237 &{\em M} &{\em  M} &{\em M} &M6    &  -       &$<$M6 I & M3 Ia\\
243 &{\em OB}&{\em OB} &-       &B2Ia  &  -       &A2 Ia & -\\
265 &{\em OB}&{\em OB} &-       &G0 Ia &  -       &F5 Ia & $<$G5 Ia \\
\hline
\end{tabular}
\end{center}
Note that classifications given in itallics have been derived from photometric 
data alone, while the 2006 classifications are derived from 2.3$\mu$m spectroscopy
(Mengel \& Tacconi-Garman \cite{mengel}, \cite{mengel08}). Crowther et al. 
(\cite{crowther}) provide a comparable table for the WR population.
\end{table*}

\begin{table*}
\begin{center}
\caption[]{Summary of the properties of the spectroscopic datasets employed in this study}
\begin{tabular}{lccccccl}
\hline
\hline
Date     & MJD & Telescope & Instrument       & Setting   & $\lambda$   & Resln. & Reference\\
         &     &      &                  &           & ($\AA$)      &       &          \\
\hline
23-25/06/01    & 52083-5     & ESO 1.52m &Boller \& Chivens & Grat. \#1  & 6000-11000 &   500  & C05 \\
         &           &                  &           &       &       &          \\
07/06/02 +&  52432  & NTT       &  EMMI            & Grat. \#7  & 6310-7835  &  2600  & C05 \\ 
06-08/06/03  & 52796-8    &      &                  & Grat.\#6   & 8225-8900  &  5000  & Negueruela \& Clark (\cite{nc}) \\
         &   &        &                  &           &       &       &          \\
12-13/06/04 & 53168-69  & VLT       & FORS longslit    & G1028z    &   7730-9480      &   $\sim$7000     & Negueruela et al. (\cite{n09})\\
         &   &          &                  & G1200R    &     5750-7310        &    $\sim$7000    & \\
           &    &       &FORS MXU          & G1200R    &     7730-9480$^a$     &   2560    & \\   
          &    &       &FORS MXU          & G1028z   &      5750-7310$^a$     &    2140    & \\
    &      &     &                  &           &       &       &          \\
25/03/05 + &53454 & VLT       &FLAMES+     & LR6        & 6438-7184  & 8600   & Ritchie et al. (\cite{benLBV})\\
29/05/05 +& 53519     &     & MEDUSA         & LR8        & 8206-9400  & 6500   & \\
13/07/05& 53564&          &                  &            &            &        &     \\
         &           &                  &           &       &       &          \\             
29-30/06/05& 53550-1 & NTT    &   SOFI        & IJ Grism     & 9500-13500 & 1000       &  Crowther et al. (\cite{crowther}) \\        
         &          &       &               & HK Grism     & 15000-22700      & 1000 & \\
         &           &                  &           &       &       &          \\             
17/02/06    & 53783 &NTT       & EMMI             &   Grat. \#6      & 5740-8730      & 1500    & Marco \& Negueruela (\cite{marco}) \\
         &           &                  &           & &      &       &          \\   
11-12/03/06& 53805-6 & VLT & ISAAC    &SWS(MR)       & 22490-23730 &  9000   & Mengel \&         \\   
        &           &                  &           &    &   &       & Tacconi-Garman (\cite{mengel})      \\ 
          &           &                  &           & &      &       &          \\   
20+29/06/08 + & 54367+76& VLT       &FLAMES+ & HR21        & 8484-9001  & 16200  & Ritchie et al. (\cite{benbin})\\
18+24/07/08 + & 54665+71     &    & MEDUSA          &        &            &            &        \\    
14+17/08/08 + & 54692+95    &     &                  &     & & & \\
04+15/09/08 +  & 54713+24     &    &                 &                    &            &         \\
19+25/09/08 + & 54728+34     &     &                 &                   &        &  \\  
14+18/05/09 + & 54965+69   &        &                &                    &        &  \\  
20/08/09  & 55063   &        &                &                    &        &  \\  
\hline
\end{tabular}
\end{center}
$^a$Note that the nominal wavelength range is given for the 2004 FORS MXU observations; in practice this can be displaced
 by a few hundred $\AA$ depending on the position of an individual star on the CCD.
\end{table*}

\begin{table*}
\begin{center}
\caption{Summary of spectroscopic data for supergiants of spectral type B2 and earlier.} 
\begin{tabular}{lcccccccccc}
\hline
\hline 
ID             &Spectral   & RA, DEC (J2000) & R & I & 1.5m   & NTT     & NTT      & VLT      &   VLT    & Notes \\  
               & Type      &                 &   &   &(2001) & (2002)  & (2003)   & (2004)    &(2008/9) &       \\
\hline 

W2a & B2Ia    & 16 46 59.7 -45 50 51.1 &14.23  &11.73 &  $\bullet$    &\#6&                 & R+z &            & {\bf A}, cand. binary, H$\alpha$ var.\\  
W6a & B0.5Iab & 16 47 04.0 -45 50 21.0 &15.80  &13.16 &  $\bullet$    &         & \#7         & R+z &  $\bullet$ & {\bf P}, cand. binary, H$\alpha$ var. \\      
W8b & B1.5Ia  & 16 47 06.0 -45 50 25.0 &-  &- &               &         &          & R+z & $\bullet$ & pulsator\\
W10 & B0.5+OB & 16 47 03.3 -45 50 34.7 &-  &- &  $\bullet$    &         &          & R+z &           & {\bf A}, binary\\   
W11 & B2 Ia   & 16 47 02.2 -45 50 47.0 &14.52  &11.91 &               &\#6+\#7&          & R+z &           & {\bf A}\\
W15 & O9Ib    & 16 47 07.0 -45 50 28.0 &16.38  &13.75 &  $\bullet$    &         & & R         & $\bullet$ & {\bf A}\\   
W19 & B1Ia    & 16 47 04.9 -45 50 59.1 &15.21  &12.37 &                &\#6+\#7& &  R+z    &           & {\bf A}, H$\alpha$ var.\\      
W21 & B0.5Ia  & 16 47 02.0 -45 51 11.0 &15.56  &12.74 &               &         &          & R+z & $\bullet$ & {\bf A}, pulsator\\
W23a& B2Ia+OB & 16 47 03.0 -45 51 08.0 &14.91  &12.07 &               &\#6+\#7&          & R+z & $\bullet$ & {\bf A}, binary, H$\alpha$ var., pulsator\\    
W24 & O9Iab   & 16 47 03.0 -45 51 11.0 &15.96  &13.24 &               &\#6+\#7&          &z & $\bullet$ & cand. pulsator\\      
W28 & B2Ia    & 16 47 04.7 -45 50 38.4 &14.26  &11.64 &               &\#6+\#7&          & R+z &           & {\bf A}, H$\alpha$ var.\\  
W29 & O9Ib    & 16 47 04.4 -45 50 39.8 &16.02  &13.38 &  $\bullet$    &         &          & R+z &           & \\  
W30a& O+O     & 16 47 05.0 -45 50 39.0 &15.80  &13.20 &               &\#6+\#7&          & R+z & $\bullet$ & binary, H$\alpha$ var. \\ 
W35 & O9Iab   & 16 47 04.2 -45 50 53.5 &16.00  &13.31 &  $\bullet$    &         &          & R+z &           & \\  
W41 & O9Iab   & 16 47 02.7 -45 50 56.9 &15.39  &12.78 &  $\bullet$    &         &          & R+z &           & \\ 
W43a& B1Ia    & 16 46 03.5 -45 50 57.3 &15.22  &12.26 &  $\bullet$    &\#6+\#7&          & R+z & $\bullet$ & {\bf A},binary, H$\alpha$ var.\\ 
W55 & B0Ia    & 16 46 59.0 -45 51 29.0 &15.25  &12.67 &  $\bullet$    &         &  &          & $\bullet$ & {\bf A}\\  
W60 & B0Iab   & 16 47 05.0 -45 51 51.0 &15.96  &13.28 &               &\#6+\#7&          & R+z & $\bullet$ & {\bf A}\\  
W61a& B0.5Ia  & 16 47 02.3 -45 51 41.6 &14.62  &12.01 &               &\#6+\#7&          &    R+z      &           & {\bf A}, H$\alpha$ var.\\
W61b& O9.5Iab & 16 47 02.6 -45 51 41.6 &16.00  &13.31 &              &\#6+\#7&          & z &           & \\  
W74 & O9.5Iab & 16 47 07.0 -45 50 10.0 &-  & - &              &         &          & R+z & $\bullet$ & {\bf A}\\
W78 & B1Ia    & 16 47 02.0 -45 49 56.0 &14.54  &12.04 &             &         &          & R+z & $\bullet$ & {\bf A}, pulsator\\
W238& B1Iab   & 16 47 05.0 -45 52 27.0 &14.98  &12.45 &  $\bullet$    &         &          & R+z & $\bullet$ & {\bf A}\\   \hline
\end{tabular}
\end{center}
{Note that W2a was also observed in 2005 March and W61a in 2005 May+July (VLT FLAMES+Medusa). Spectral classifications are from Negueruela et al. (\cite{n09}).
The final column indicates whether a given star
is identified as a {\bf P}eriodic or  {\bf A}periodic photometric variable, (candidate) binary, H$\alpha$ variable and/or (candidate) photospheric pulsator (Sect. 3).}
\end{table*}

\subsection{Photometric data}
Comparative analysis of multiepoch  photometric data obtained prior to 2001  was not attempted for the 
above reasons and, with the exception of West87,  the limited number of stars observed in such works, which 
precluded the determination of empirical offsets between different datasets (e.g. Fig.1)\footnote{Note
that the limited stellar identifications and lack of  co-ordinates presented by Piatti et al. (\cite{piatti}) 
 limit the utility of those observations.}. Hence the only  robust results  to be drawn from these 
observations are the coarse spectral classifications derived from  
colour indices for the (contemporaneous) datasets presented by  West87, Lockwood (\cite{lockwood74}), 
Koornneef (\cite{koor}) and Piatti et al. (\cite{piatti}). Unfortunately these just permit the classification 
of stars into OB and M supergiant classifications, for which we found the former to contain stars of spectral 
type from late O to mid GIa$^+$. These results are summarised in Table 2.

These problems persist for modern, post-2001 datasets. This is apparent in the systematic 
offset between the V band datasets present in C05 and Bonanos (\cite{bonanos}; Fig. 1) due to the difficulty 
in determining the zero point offset because of the lack of appropiate red standards\footnote{Bonanos 
(\cite{bonanos}) successfully made such  a correction for R and I band, but the methodology employed - 
comparison of the brightest  100 stars in common between the 2 datasets - was not completely successful for the V band data.
We suspect this to be 
because the brightest stars in the V band were dominated by foreground stars of relatively low reddening 
in comparison to the stars within Wd1.}. However the large number of stars in common between West87, C05 and 
Bonanos (\cite{bonanos}) enabled the offsets to be determined empirically and hence a direct comparison to be 
made between the three epochs (Fig. 1). 

This approach allows potentially variable outliers to be identified, albeit with three corrolaries. Firstly 
observations of a particular  star are  not always present in  each dataset; for example W4, 20, 26, 32, 42, 75 \& 
243 are absent from C05. Secondly, observations of individual objects are likewise not present in all 3 
wavebands of a given dataset. For example, comparable  photometry  for W243 between West 87 and 
Bonanos (\cite{bonanos})  is only available in the V band, thus the apparent discrepancy  between the 
two epochs is impossible to verify\footnote{The same is  the case for W44 \& 57a, and W57a  in the comparison 
between West87 and C05 datasets.} (complicated by the fact that the empirical offset  is ill defined for 
stars of this magnitude). Finally, issues arising from  filter and photographic emulsion response to
very red objects (West87, Bessell \cite{bessell}) cast doubt upon the reality of the offsets between 
the data presented by West87 for the   RSGs W26 \& 237 and other epochs,  although this would {\em not} 
explain the apparent  variability  of W237 in V \& R bands between 2001-6 which we consider likely to be
 of astrophysical origin (Sect. 7). 

Given the above, in addition to W237 we may only tentatively identify a handful of stars - W9, 57a \& 243 -  
as {\em potentially} variable, although with only 3 epochs 
of observations over a 20~yr baseline and a lower limit on detectable variations of $>$0.5mag (based on the 
scatter between datasets evident in Fig. 1),  these are hardly stringent constraints. Nevertheless,
 foreshadowing the detailed results and discussion presented in  Sects. 3-7, both hot and cool transitional 
stars may demonstrate long term ($>$decades) variability of many magnitudes amplitude; we would  
 have expected to identify such behaviour photometrically if present within the evolved stellar 
population of Wd1.

\subsection{Spectroscopic data}

Prior to the discovery of a rich population of WRs  within Wd1, the only spectroscopic survey to have been 
undertaken was that of West87. Motivated by this discovery, several optical and near-IR studies have 
subsequently  been made (Table 1); the {\em published} results of all these works are summarised in Table 2.
 Given the varied science goals of the proposals from which these observations have been drawn, 
the resulting spectra sample a number of different wavelength ranges  and corresponding resolutions; these 
are summarised in Table 3 with, where available,   a reference for a more detailed description including 
the reduction  procedure adopted.
 Of these data, those obtained in 2003 June (NTT/EMMI), 2005 March, May \& July (VLT/FLAMES), 
2006 February (NTT/EMMI) and  2008 September \& 2009  May \& August (VLT/FLAMES) are published  here for the first time, as 
are the observations of the YHGs from 2004 June (VLT/FORS) with   the corresponding data for the  OB stars 
 found in Negueruela et al. (\cite{n09}). In total this amounts to in excess of a hundred new spectra. 

Special mention must be made of the optical  spectra obtained in 2001 June (C05) and the near-IR spectra 
obtained in 2006 March (Mengel \& Tacconi-Garman \cite{mengel}, \cite{mengel08}). The former are
 of significantly lower resolution and S/N than  later epochs.
Following the analysis presented in Negueruela et al. (\cite{n09}) 
 we may utilise these data to adequately distinguish between
 extreme B hypergiants/WNL stars (Group 1 stars following C05, with the additional criteria that H$\alpha$ must be in emission),  late O/early B supergiants 
(Group 2; O9-B2 I),  mid-late B super-/hypergiants (Group 3; B2.5-9I$^{(+)}$) and YHGs. Thus, while these data 
are inadequate for the analysis of line profile
 variability\footnote{However they are sufficient to identify  dramatic changes in line profiles 
such as the appearance of  H$\alpha$ emission in  W23a and W43a; C05 and Sect. 3.} 
they do permit the gross spectral classification of cluster members, 
hence extending the baseline   for the identification  of e.g. the characteristic temperature excursions of LBVs
 by a further year.

The latter spectra sample a relatively poorly studied region of the near-IR waveband, for which a comprehensive spectral atlas 
has yet to be constructed. Nevertheless, the relative strength of the CO bandheads may permit classification of cool 
supergiants (Figer et al. \cite{figer06},  Yamamuro et al. \cite{yama}). Mengel \& Tacconi-Garman (\cite{mengel},\cite{mengel08}) use this 
diagnostic to classify the 4 RSGs within Wd1, while  the lack of CO bandhead absorption features in the spectra of 
W4, 8a, 9, 12, 32 \& 265  suggest  a spectral type of G5 Ia or earlier.

Summaries of the data available for late O/early B supergiants and the more evolved,  transitional stars are presented in 
Tables 4 \& 5  respectively. A comparable table for the WRs may be found in  Crowther et al. (\cite{crowther}), noting that 
the three WRs included in Table 5 are the sole objects for which multiple epochs of high resolution data are available, thus 
permitting analysis of line profile evolution. Given the recent 
comprehensive analysis by Ritchie et al. (\cite{benLBV}), the LBV W243 has been  omitted from this study. 


\begin{landscape}
\begin{table}
\caption{Summary of the  spectroscopic datasets available for B2 and later super-/hypergiants and Wolf-Rayets.}
\label{target}
\begin{tabular}{l|c|c|cc|ccccccccc|l}
ID             &Spect. Type  & RA, DEC (J2000) & R & I  
& 06/01    & 06/02  & 06/03     & 06/04     &  03/05    &  05+07/05 & 02/06   & 03/06 & 2008/9    & Notes \\
\hline
\hline
&&&&&&&\\
W4             & F3 Ia$^{+}$    & 16 47 01.42,~-45 50 37.1 &11.18 &9.15 
&$\bullet$ &\#6+\#7 & \#6  &           & LR6+8 & LR6+8 &         & $\bullet$  &            
&\textbf{A,P+R} \\
W5 (S)         & WNL/BIa$^{+}$ & 16 46 02.97,~-45 50 19.5 & 14.98 &12.48 
&$\bullet$ &        &  \#7     & R+z& LR8 &           &         &      &        
&\textbf{A}\\
W7             & B5 Ia$^{+}$    & 16 46 03.62,~-45 50 14.2 &12.73 &9.99 
&$\bullet$ &\#6+\#7 & \#6+\#7  &  &           & LR6+8 &         &       &       
&\textbf{A}\\
W8a            & F8 Ia$^{+}$    & 16 47 04.79,~-45 50 24.9 & 12.64&9.89 
&$\bullet$ &\#6+\#7     &           &           &           & LR6+8 &    &$\bullet$     &               
&\textbf{A}\\
W9             & sgB[e]        & 16 47 04.14,~-45 50 31.1 & 14.47&11.74
&$\bullet$ &\#6+\#7     & \#7       & R+z &           &           &     &$\bullet$    &               
&\textbf{A,X,P+R}\\
W12a           & F1 Ia$^{+}$    & 16 47 02.21,~-45 50 58.8 & 13.54&10.42 
&$\bullet$ &\#6+\#7 & \#6   & R+z &           &           &   &$\bullet$       &              
&\textbf{A,R}          \\
W13            & WNVL/BIa$^{+}$+OB   & 16 47 06.45,~-45 50 26.0 & 14.63&12.06 
&$\bullet$ &\#6+\#7     &          & R+z  & LR8 & LR6+8 &  $\bullet$& &$\bullet$    
&\textbf{X,E}, binary\\
W16a           & A5 Ia$^{+}$    & 16 47 06.61,~-45 50 42.1 & 12.82&9.90 
&$\bullet$ &\#6+\#7 &\#6+\#7   & z& LR6+8 &           &$\bullet$& $\bullet$ &             
&\textbf{A,P}\\
W20            & M5 Ia      & 16 47 04.70,~-45 51 23.8 & - & - 
&$\bullet$ & \#6    &          &  &       &           &    &$\bullet$     &             
&\textbf{A,R}\\
W26            & M5-6 Ia      & 16 47 05.40,~-45 50 36.5 & 12.63 & 9.19 
&$\bullet$ & \#6    & \#7      &  &       &           &$\bullet$&    $\bullet$ &         
&\textbf{R}\\
W32            & F5 Ia$^+$       & 16 47 03.67,~-45 50 43.5 &  -    & - 
&$\bullet$ & \#6+\#7    &       &  &       &           &   &$\bullet$ &             
&\textbf{P}\\
W33            & B5 Ia$^{+}$    & 16 47 04.12,~-45 50 48.3 &12.78 & 10.04
&$\bullet$ &\#6+\#7 &           & R+z & LR8 &           &         &   &          
&\textbf{A}\\
W42a           & B9 Ia$^{+}$    & 16 47 03.25,~-45 50 52.1 & -& -
&$\bullet$ &        &           & R+z &           & LR6+8 &         &      &        
&\textbf{A}\\
W44 (L)        & WN9           & 16 46 04.20,~-45 51 06.9 & 15.61&12.52 
&$\bullet$ &  \#6+\#7      &       & R+z & LR8 & LR6 &         &        &      
&\textbf{A,P} \\
W57a           & B4 Ia          & 16 47 01.35,~-45 51 45.6 &13.83 &11.13 
&          &\#6+\#7     &           & R+z& &           &    &     &$\bullet$    
&\textbf{A}, cand. binary\\
W70            & B3 Ia          & 16 47 09.36,~-45 50 49.6 & 14.10 &11.29 
&$\bullet$ &\#6+\#7 &           & R+z &           & LR6 &$\bullet$& &               
&\textbf{A}\\
W71            & B2.5 Ia        & 16 47 08.44,~-45 50 49.3 &14.06 &11.16 
&$\bullet$ &\#6+\#7 &           & R+z &  &           &     &    &$\bullet$   
&\textbf{A}, cand. binary \\
W75            & M4 Ia        & 16 47 08.93,~-45 49 58.4 & - & - 
&  &  &           &   &  &           & $\bullet$    & $\bullet$    &   
&\textbf{P} \\
W237            & M3 Ia      & 16 47 03.09,~-45 52 18.8 & 12.63 & 9.19 
&$\bullet$ &      &             &  z & LR6+8   & LR6+8          &       & $\bullet$        &
&\textbf{A,R}\\
W239(F)        & WC9           & 16 47 05.21,~-45 52 25.0 & 15.39 & 12.90
&$\bullet$ &        &           &           &     &      &           &         & $\bullet$
&\textbf{A,X,P}, binary \\ 
W241(E)        & WC9           & 16 47 06.06,~-45 52 08.3 & -    & -
&$\bullet$ &        &           &           & LR8 & LR6+8&           &         &    
&\textbf{A,X}, cand. binary \\
W265           & F1-5 Ia$^{+}$    & 16 47 06.26,~-45 49 23.7 &13.62 &10.54 
&$\bullet$ &\#6+\#7 & \#7   & R+z &           & LR6+8 &         &$\bullet$ & $\bullet$
&\textbf{A,R} \\
\hline
\end{tabular}
Note the spectral classification of the OB stars  are from Negueruela et al. (\cite{n09}), the YHGs from this work (Sect. 6) and the
RSGs from Mengel \& Tacconi-Garman (\cite{mengel}). 
 Where appropriate, the 
 Notes column lists stars as {\bf E}clipsing or {\bf A}periodic photometric variables (Bonanos \cite{bonanos}),  
{\bf X}-ray (Clark et al. \cite{c08}) and  {\bf P}oint and/or {\bf R}esolved radio (Dougherty et al. \cite{d09}) sources. 
Finally, the binary status of 5 stars - as determined
 by Ritchie et al. (\cite{benbin}) from multiepoch radial velocity spectroscopic data - is presented, noting that 
insufficent data currently exist to confirm the status of the other stars (but note Sect. 6 for W265). 
\end{table}
\end{landscape}

\begin{figure}
\begin{center}
\resizebox{\hsize}{!}{\includegraphics{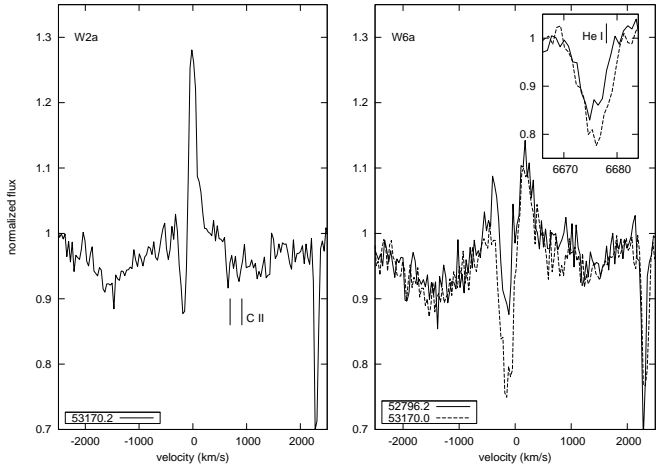}}
\caption{  Left panel: H$\alpha$ for  W2a (B2 Ia) from 2004 - the line is strongly in emission in this 
spectrum but was absent in the low S/N spectrum from 2001 (not shown). Right panel:  H$\alpha$ profile for 
 W6a (B0.5 Iab) in 2003-4; note the temperature dependent He\,{\sc i}${\lambda}{\lambda}$ 6678, 7065{\AA} lines also appear to  vary in strength.}  
\end{center}
\end{figure}

\begin{figure}
\begin{center}
\resizebox{\hsize}{!}{\includegraphics{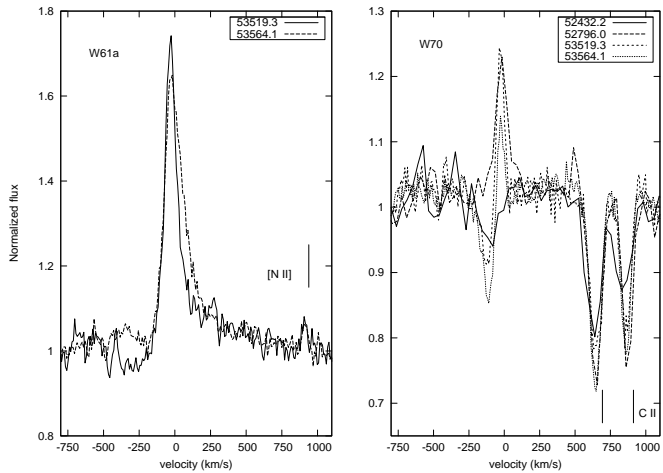}}
\caption{LPV in H$\alpha$ for  W61a (B0.5 Ia)  and W70 (B3Ia). Note the variability in the strength of the temperature
dependent C\,{\sc ii} photospheric lines in W70.}

\end{center}
\end{figure}

\begin{figure}
\begin{center}
\resizebox{\hsize}{!}{\includegraphics{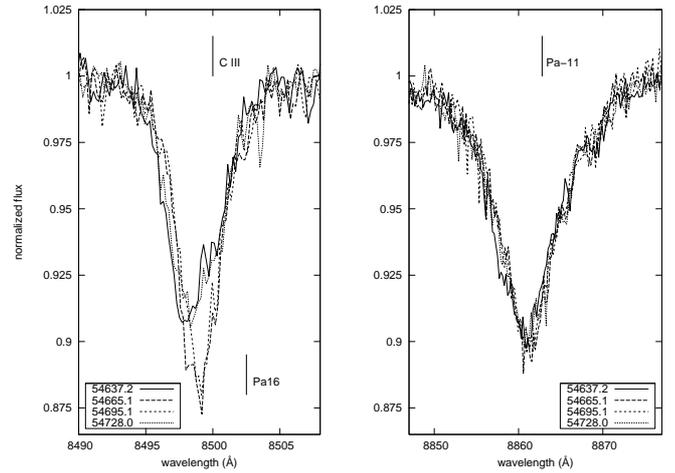}}
\caption{LPV in the 2008-9 VLT/FLAMES spectra C\,{\sc iii}+Pa16 blend in W24; in comparison little variability is observed in the unblended
Pa11 profile. Note that the  strength of C\,{\sc iii} line in the 2002 NTT and 2004 VLT spectrum, which are not shown for 
clarity, are comparable to the strong (e.g. MJD 54665.1) and weak (e.g. MJD 54637.2) `states' respectively.}
\end{center}
\end{figure}

\begin{figure}
\begin{center}
\resizebox{\hsize}{!}{\includegraphics{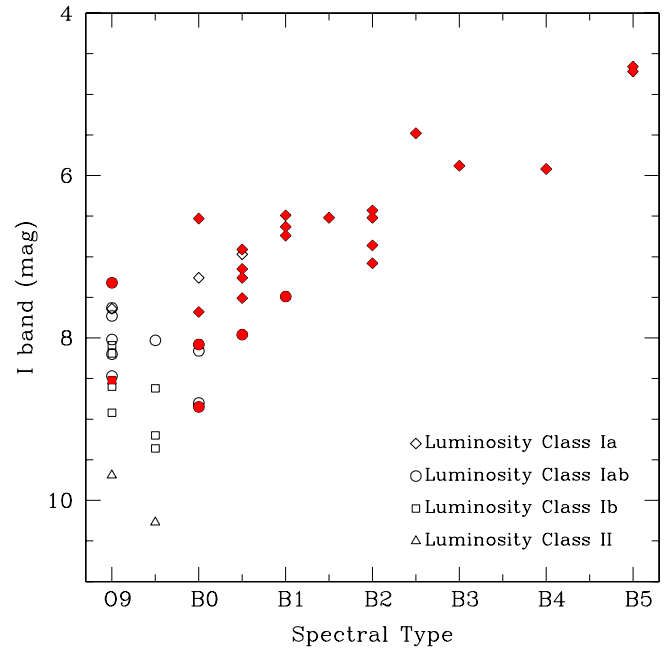}}
\caption{Plot of I band magnitude versus spectral type for the late O - mid B stars within Wd1 (stars individually 
dereddened according to the prescription given in Negueruela et al. \cite{n09}). The  
photometric variables identified by   Bonanos (\cite{bonanos}) are indicated by the filled symbols.} 
\end{center}
\end{figure}

\begin{figure}
\begin{center}
\resizebox{\hsize}{!}{\includegraphics{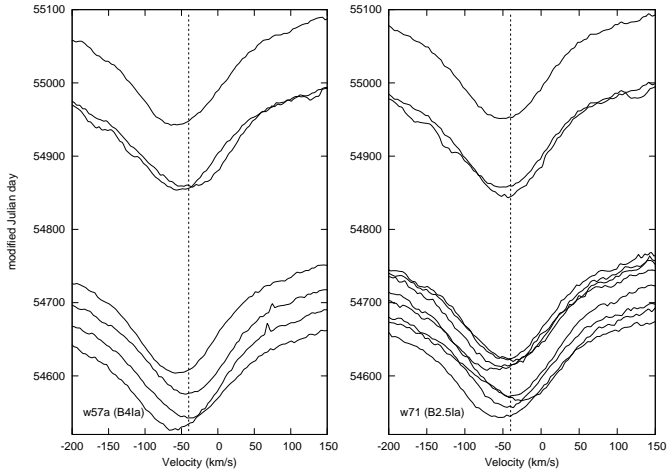}}
\caption{LPV in  Pa11 indicative of photospheric pulsations  for the BHGs W57a (B4 Ia) and
W71 (B2.5 Ia). Similar variability is also present in the N\,{\sc i} lines.}
\end{center}
\end{figure}

\begin{figure}
\begin{center}
\resizebox{\hsize}{!}{\includegraphics{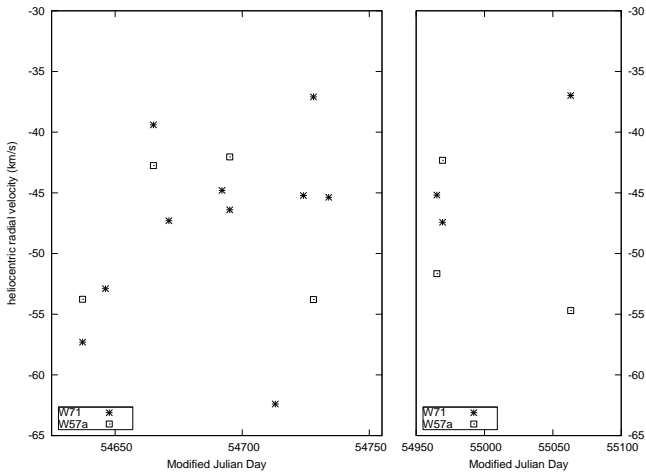}}
\caption{Plot of RV variations versus time  for the Pa11 line in the BHGs W57a (B4 Ia) and
W71 (B2.5 Ia). The RV is the error-weighted
average of the Pa-11..14 lines,  with an uncertainty of  $\pm$3~kms$^{-1}$.} 
\end{center}
\end{figure}

\begin{figure}
\begin{center}
\resizebox{\hsize}{!}{\includegraphics{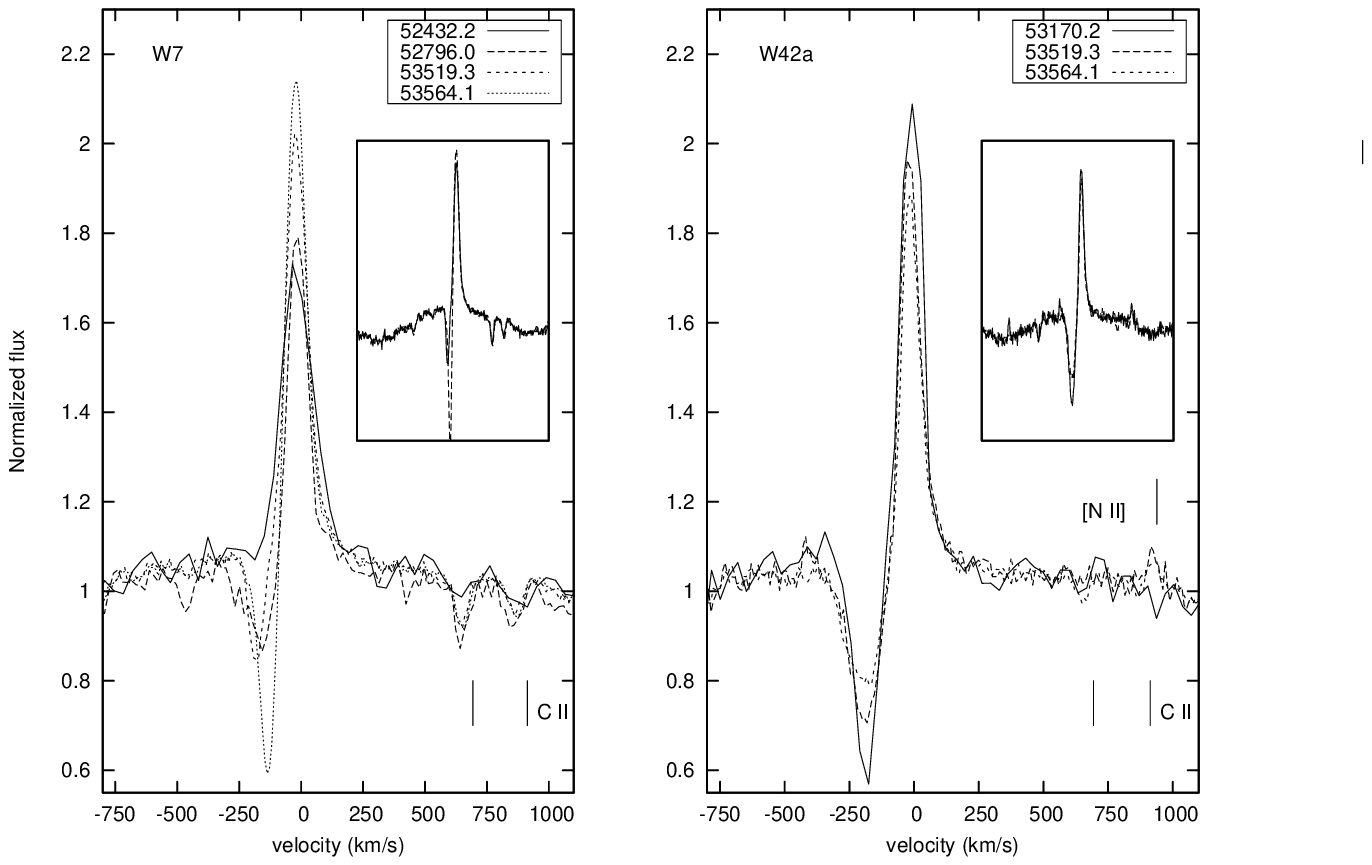}}
\caption{LPV in the H$\alpha$ line of W7 (B5 Ia$^+$)   and W42a (B9 Ia$^+$). The insets
show the Halpha profile with a log flux scale to highlight the broad emission wings.}
\end{center}
\end{figure}

\begin{figure}
\begin{center}
\resizebox{\hsize}{!}{\includegraphics{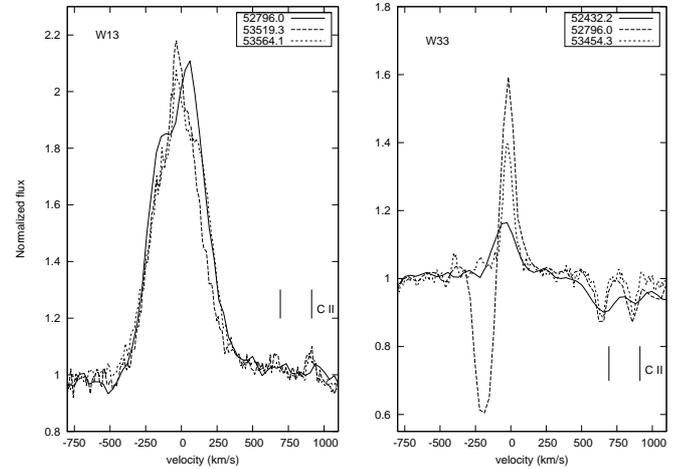}}
\caption{LPV in the H$\alpha$ for W13 (WNVL/B Ia$^+$+OB) and W33 (B5 Ia$^+$).}
\end{center}
\end{figure}

\begin{figure}
\begin{center}
\resizebox{\hsize}{!}{\includegraphics{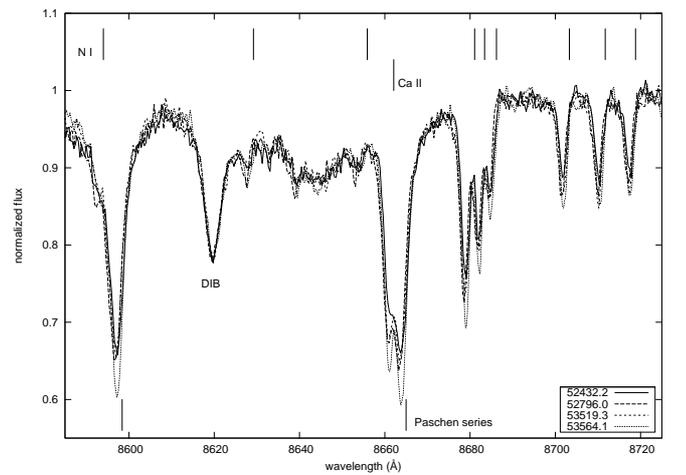}}
\caption{Pulsational variability in the Paschen series and low excitation metallic lines of W7 over timescales as short as 35 days.}
\end{center}
\end{figure}

\section{The O9-B2 supergiants}

Prior to 2008, comparatively  few  spectroscopic data were  available for individual O9-B2 
supergiants (Table 4), restricting analysis primarily to the detection of line profile variability (LPV) 
in the H$\alpha$ line. Of the stars  with multiple spectra covering this transition, 12 have two or more 
 observations for which R$\geq$2600, while  a further 6 have two epochs of data of which one is of low resolution (R$\sim$500) 
and S/N.
 Of these 18, 9 stars demonstrated significant H$\alpha$ LPV. This is illustated  for W2a, 6a and 61a in Figs. 2-3\footnote{Note that throughout the paper the {\em rest} wavelengths of transitions are indicated in Figures, while the stellar 
spectra demonstrate a systemic blueshift of $\sim$40kms$^{-1}$.}, 
while it is apparent for  W19, 23a, 28, 30a and 43a  by comparison of 
 the spectra presented in C05, Clark et al. (\cite{c08}) 
and  Negueruela et al. (\cite{n09}); the  spectra of the final example - W28 - are omitted due to their comparatively low S/N.  
Poor temporal sampling  places few constraints on the timescales of the variability, but  in the case 
of W61a significant changes occur over  a few tens of days (Fig. 3).
Morel et al. (\cite{morel}) monitored 12 O9-B2 Ia-Ib field supergiants, finding  conspicuous variability 
on comparably rapid timescale in many of the stars with H$\alpha$ profiles 
dominated by wind emission, including -  as found for W19 -   a  transition from  absorption to full emission in 
HD 37128 (B0 Ia).  Moreover all of the H$\alpha$ line profile morphologies  demonstrated by the OB supergiant
 population within Wd1  were found to have identical counterparts in the  samples 
 presented in Kaufer et al. (\cite{kaufer})
 and Morel et al. (\cite{morel}).  
 
As such the  variability of the Wd1 members does not appear atypical for late O/early B supergiants, while the poor sampling frequency of the current 
dataset suggests that many more stars are likely to be intrinsically variable. Such a supposition is supported by all 9
stars displaying H$\alpha$ LPV  being drawn from the subset of stars with 
2 or more epochs of high resolution data.
Although  is not expected that such stars will have evolved far enough to encounter the LBV phase, the physical origin 
of such variability is currently uncertain, with pulsational, magnetic and/or binary modulation of the wind all possible. 
In this respect it is of interest  that W2a, 6a, 23a, 30a and 43a have all been identified as binaries (Clark et al. \cite{c08}, 
 Negueruela et al. \cite{n09} and Ritchie et al. \cite{benbin}).

Of the 9 stars with both NTT (2002) and VLT (2004) I band spectra, only one star - W24 - showed evidence for variability; intriguingly 
this was   most pronounced in the C\,{\sc iii}+Pa16  8500{\AA} blend rather than in other stronger,  unblended Paschen
 lines. Motivated by this we examined the full VLT/FLAMES 2008-9 spectroscopic dataset (with analysis of the comparable
 datasets for the full sample presented in a future dedicated paper). These data confirmed that the blend varied in 
both strength and radial velocity (RV) of the profile centroid over short  timescales ($\sim$24d; Fig. 4). Given the 
weakness of the adjacent Pa 16 transition  and  Pa15+Ca\,{\sc ii} \& Pa13+Ca\,{\sc ii}  blends as well as  
the lack of comparable LPV in the Pa11 line  we associate this behaviour  with the C\,{\sc iii} 8500{\AA} line 
rather than  Pa16 or Ca\,{\sc ii}$\lambda$ 8498.
The physical origin  of this behaviour remains uncertain, with no periodicity or systematic secular evolution 
 present in the current data. Ritchie et al. (\cite{benbin}) report similar  LPV  for the Paschen series in the 
OB supergiants W8b, 21, 23a and 78, which they attribute to photospheric pulsations; it is tempting
to include  W24 in this category, although why pulsations should solely be visible in  C\,{\sc iii} 8500{\AA} and not in the 
Paschen series   is unclear. 

Another group of  spectroscopically variable stars are the rare Of?p objects, in which the 
phenomenon {\em may} be  linked to the presence of strong  magnetic fields (e.g.  Naz\'e et al \cite{naze}). 
The defining feature of these  stars  is  (variable) emission in the  C\,{\sc iii}${\lambda}{\lambda}$~4647,50,51 multiplet\footnote{Note that the 
 C\,{\sc iii}  8500{\AA} line is the singlet counterpart (3s$^1$S-3p$^1$P$^0$) to this multiplet.}. Unfortunately this feature is currently inaccessible  in W24, 
and while it is a 
hard  X-ray source it  is not over luminous for its bolometric luminosity as might be expected for an Of?p star 
(Clark et al. \cite{c08}). Nevertheless, given the presence of a magnetar within Wd1 (Muno et al. \cite{muno}) and 
the suggestion that these are the descendents of  highly magnetised  progenitors, the possibility that W24 could be a member of such a population is 
intriguing and warrants further investigation.

Regarding variability amongst the OB supergiant population within Wd1, we may also utilise the results of Negueruela et al. (\cite{n09})  to reinterpret the findings of Bonanos (\cite{bonanos}). In Fig. 5 we plot a semi-empirical HR diagram for all the OB supergiants within Wd1 for which a spectral classification is 
available,  highlighting the location of the photometric variables identified in the latter work. This emphasises that
 such stars are predominantly  concentrated amongst the more luminous, evolved  SGs; a similar result to that  found by 
Fullerton et al. (\cite{fullerton}). Given the apparent dependence on 
evolutionary state we assume that the 
{\em photometric} variability  is most likely intrinsic to the star and hence not binary in origin. 

It is tempting to associate this 
behaviour with  the $\alpha$ Cygni variables, of which the subset of late O/early B supergiants 
demonstrate quasi-periodic oscillations with $\leq$0.1mag 
amplitude  (e.g. van Leeuwen et al. \cite{vL}). Unfortunately, the observations of Bonanos 
(\cite{bonanos}) -  comprising two 10 day blocks separated by 20 days - are ill-suited to identifying the photometric 
modulation of $\alpha$ Cygni variables, which characteristically occur  over  2-6 week timescales for early OB 
supergiants, extending to  hundreds of days for cooler super-/hypergiants (van Leeuwen et al. \cite{vL}).

Finally, four stars - W6a, 10, 11 and 28 - were also observed by West87; given their spectral classifications
 agree to  $\leq$2 subtypes we find no evidence for their long term secular evolution. Nevertheless, of the 23 stars
 considered here,  we find that  all but four are either spectroscopic or photometric variables (Table 4). 

\section{The mid-late B super-/hypergiants}

The 6  stars with spectral types B2.5-9 naturally divide into two subgroups. The first, containing W57a, 70 \& 71, are 
earlier (B2.5-4 Ia) and have comparatively weak H$\alpha$ emission, while the latter, comprising W7, 33 \& 42a, are later 
(B5-9 Ia$^+$), and  demonstrate stronger H$\alpha$ emission with  broad wings (likely due to electron scattering).  
Despite this, all were found to be    spectroscopically variable. 

Of the B2.5-4 Ia stars,  the H$\alpha$  profile of W70 was observed in absorption  in 2002, full emission in 2003 
and P Cygni emission in 2005 (Fig. 3). The C\,{\sc ii}${\lambda}{\lambda}$6578, 6582 transitions 
also appeared to vary in both strength and the RV of the line  centroid between observations, with changes in the  profiles 
of both species observed over the shortest timescale probed (45~d)\footnote{While we regard the detection of RV changes as 
provisional, given the relatively low S/N and resolution of the observations, the changes in the line strength are robust, 
amounting to $\sim$15\% in Equivalent Width between the identical resolution 2005 FLAMES observations.}. 
In  comparison, the early B supergiants HD 14134 (B3Ia) and HD 43384 
(B3Iab) also showed a rapid ($\leq14$~d) transition from H$\alpha$ absorption to emission (Morel et al. \cite{morel}), 
although neither they nor the other B2-4 supergiants from 
that study demonstrated variability in the C\,{\sc ii} lines. These transitions are temperature dependent and are  absent in 
supergiants earlier than  B0.7 and later than B9 and peak at B3, although their apparent abundance  dependance
have precluded their use as an accurate diagnostic (Negueruela et al.  \cite{n09}). Hence 
we may conclude that both wind and photosphere are variable, although we may not constrain the magnitude of the 
temperature excursions exhibited by the  latter.

W71 also demonstrated significant LPV in H$\alpha$, which was  present in emission in 2002 but absent in 2004, while it
 was absent in both spectra of W57a. For both objects the S/N and resolution of the spectra were too low to 
comment on the behaviour of the nearby C\,{\sc ii} and  He\,{\sc i} features and for that reason are not shown here. 
Included in the binary survey (Ritchie et al. \cite{benbin}), the multiple I band observations of both stars demonstrated 
rapid  RV variability in the Paschen series and N\,{\sc i} lines (Figs. 6-7). Building on these results we found 
$\Delta$RV$\sim$12kms$^{-1}$ and 25kms$^{-1}$ for W57a and 71 respectively with no period apparent  in either dataset;
 consistent with  random sampling of a short (pulsational) {\em quasi}-period,  comparable in both respects to the 
photometric pulsations observed in other early and late B 
supergiants (Kaufer et al. \cite{kaufer97}, \cite{kaufer06}). The change in strength of the 
N\,{\sc i} lines in both stars indicates that pulsations were present, being inconsistent with an origin for the 
RV changes in binarity alone. These imply moderate variations in the spectral type corresponding to 
$\Delta$T$\sim$2000K,   noting that modeling 
of the non-radial pulsations of the B supergiant star HD64760 revealed  a comparable difference between 
minimum and maximum surface temperature (3500K; Kaufer et al. \cite{kaufer06}).

The cool B hypergiants W7, 33 and 42a also demonstrated pronounced H$\alpha$ LPV,
although an emission component was always observed (Figs. 8-9). The C\,{\sc ii} doublet was also 
found to be variable in W7; this behaviour was  mirrored in 
 the Paschen series and low excitation metallic lines present in the I band spectra (Fig. 10).  
Variations in  both the strength and  RV of the  C\,{\sc ii} doublet have previously been observed
 for the B8 Iae star  HD 199478 (Markova et al. \cite{markova}); the similarity of the resultant 
spectroscopic period to the photometric period providing strong evidence  for photospheric pulsations in this star. 
Identical LPV is also present in the I band spectra of  W42a (not shown for brevity); in conjunction with 
comparable behaviour in both hotter and cooler supergiants within Wd1 (W70 and 265 respectively) we 
conclude that both W7 and W42a are also pulsating. As with W57a and 71, the range
of wavelengths  sampled by these spectra do not include particularly sensitive temperature diagnostics, 
although changes in spectral type by $\pm$1 subtype may be inferred for both W7 and W42 over timescales of 
$\sim$35 days (Fig. 10).

However, as with the  B2.5-4 Ia supergiants, the physical origin of the H$\alpha$ variability is unclear.
Comparable P Cygni profiles and LPV in H$\alpha$ were  found   for HD 199478,   HD 92207 (A0 Ia; Kaufer et al. 
\cite{kaufer}), and the LBV  HD 160529 (Stahl et al. \cite{stahl03}).  Both HD 199478 and  
HD 92207 undergo short lived (20-60~d) episodes of enhanced blueshifted absorption leading to the appearance 
of a P Cygni profile (Kaufer et al. \cite{kaufer}, Markova et al. \cite{markova}); a timescale comparable to that 
inferred for the LPV in our stars ($\leq$45 days). Such `high velocity absorption' (HVA) events are assumed to be 
due to a transient  increase in the  optical depth along the line of sight in an aspherical  wind, 
although the origin of the asphericity is unspecified. However, in W7, 33 and 42a the increase in  strength 
of the P Cygni absorption trough was accompanied by an increase in the strength of the emission component of
 the line. Such correlated behaviour is not observed in HD 199478 and  HD 92207, but is seen in  the  LBV  
HD 160529 (Stahl et al. \cite{stahl03}), where it is instead attributed to 
a long term secular change in the {\em global} properties of the stellar wind such as mass loss rate and 
terminal velocity. Further  observations are clearly required to distinguish between both possibilities for the B 
hypergiants within Wd1

In summary, evidence for both photospheric (pulsational) and wind variability are found for both sub-groups of B 
super-/hypergiants, although no causal link between both phenomena may be drawn from the current data nor physical 
cause for wind variability inferred. However, with regard to this  we note that no binary markers are found 
for any of these stars;  likewise there is no unambiguous evidence for either long term  secular evolution  or  
canonical  LBV excursions over the 5-9yr observational baseline for individual stars 
(Table 5)\footnote{I band spectra  for LBVs in both high and low temperature phases are provided by 
Ritchie et al. (\cite{benLBV}) and Munari et al. (\cite{munari}), while observations of both early and 
late B/early A supergiants - the range over which a putative  LBV within Wd1 might be expected to evolve - 
 are presented in C05 and Munari \& Tomasella (\cite{MT}). These data clearly demonstrate that changes 
in the spectral morphology 
resulting from a canonical LBV excursion would  be identifiable in our 
data.}. Such a result is not unprecedented; P Cygni has demonstrated only weak secular evolution 
since  its outburst in 17th Century  - which would have been undetectable in our data - with 
 strong  ($>$1 mag.) LBV excursions also absent (Lamers \& de Groot \cite{lamers}, de Groot et al. \cite{deG}). Likewise 
the B hypergiants VI Cyg 12 and $\zeta^1$ Sco - which show a remarkable spectral similarity to the BHGs within Wd1 -
 have also been photometrically stable for a century  (Clark et al. \cite{c10}, Sterken et al. \cite{sterken}) 

In relation to this,  the spectral types determined by West87 and C05 for W7, 42a, 70 \& 71  are 
discrepant (Table 2).  However given the difficulties of I band  classification (Negueruela et al. \cite{n09}) we
 strongly suspect these arise from the low S/N and resolution of the spectra presented by West87 rather than from 
any intrinsic long term variability. In conjunction with photometric classifications (Table 2) we may therefore also 
exclude any long term transition to a  significantly higher  or lower temperature for these stars, such as the decades
 long excursions seen  for  R127 and  \object{M33 Var A}  (Walborn et al. \cite{walborn}, Humphreys et al. \cite{m33};
 Sect. 6).

The final OB supergiant star we turn to is the sgB[e] star W9. Given their co-location in the HR diagram, there has been 
much speculation on the evolutionary links between sgB[e] stars and LBVs. Both photometric (e.g. Bergner et al. 
\cite{bergner},  Zickgraf et al. \cite{z96}) and spectroscopic (e.g. Zickgraf \cite{z03}) variability have been observed
 in these stars but to date these have not been unambiguously associated with characteristic LBV excursions or eruptions.
In this respect W9 is of interest, given the evidence for changes in its unexpectedly high mass loss rate (Dougherty et
al. \cite{d09}) and the detections of short term aperiodic photometric fluctuations (Bonanos \cite{bonanos}). A full 
description of the optical and IR spectroscopic dataset is presented in Ritchie et al. (submitted) but,
perhaps unexpectedly, we found no evidence for pronounced spectroscopic variability over the 
course of the observations (2001-4; noting the comparatively low resolution data would not resolve LPV similar 
to that  observed in other hot supergiants).

\begin{figure}
\begin{center}
\resizebox{\hsize}{!}{\includegraphics{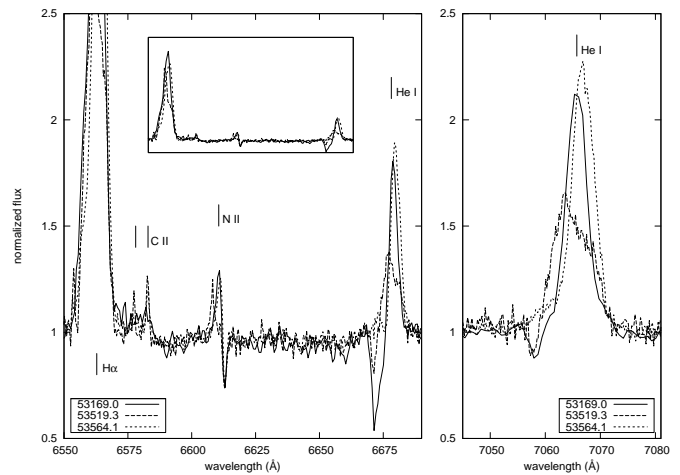}}
\caption{LPV in H$\alpha$ and He\,{\sc i} transitions for the candidate WN9 binary W44. The inset in the left panel
shows the same wavelength range but rescaled in flux to show the whole H$\alpha$ profile. }
\end{center}
\end{figure}

\section{The Wolf Rayets}

 WRs have long been  implicated in the LBV phenomenon; in 
their hot, quiescent state LBVs such as \object{AG Carinae} display a WN9-11 spectral  morphology, while the WN3 
star  HD~5980 underwent  a major eruption resulting in a transition to significantly cooler 
spectral types  in the 1980s (WN6 and B1.5Ia$^+$ ; Koenigsberger et al.  \cite{koen}).
Comparison of the optical  and near-IR spectroscopic 
datasets  (Clark \& Negueruela \cite{c02}, Negueruela \& Clark \cite{nc}
Crowther et al. \cite{crowther}) indicate that no such extreme behaviour is  evident for any of the WRs within Wd1
between 2001-5, with 
 the discrepancies between the subtypes  of individual objects being attributable to a combination of low S/N and resolution 
spectra and a poorly callibrated I band classification scheme. Unfortunately, the faintness of the WR 
population limits the utility of the long term photometry; of the  8 stars observed on 
two or more epochs\footnote{WRs A, B, E, F, G, L, M \& S}, only W44 (=WR L) is 
potentially  variable.

As with OB supergiants, line profile  variability has  been observed  in a large number of both WN and WC WRs
 (e.g. Robert \cite{robert}, Lepine \& Moffat \cite{lepine}), typically taking the form of a  superposition of variable 
subpeaks on an underlying broad emission line and attributed to both small (stochastic clumping) 
and large scale (co-rotating) wind  structure. The origin of the latter is uncertain, with both photospheric and binary origins 
advanced for the quasi-periodic line profile variability observed in e.g. \object{EZ CMa} (=WR6; Morel et al. \cite{morel}, Georgiev et 
al. \cite{georgiev}), although apparently unambiguous binary modulation  has been reported in a number of WR colliding wind 
systems  by Stevens \& Howarth (\cite{stevens}).

Including the extreme B hypergiants W5 (=WR S) and W13, which  show a close continuity in 
physical properties to the WN9 star W44, a total of five WRs have multiepoch {\em optical} spectroscopic data, 
the others are the WC9 stars W239 (=WR F) and W241 (=WR E). Negueruela et al. (\cite{n09}) present the 
highest resolution and S/N  I band spectrum of W5 available to us; no evidence for long term  line profile 
variability between  2001-4 was visible in these data. W13 is discussed in depth in  Ritchie et al. 
(\cite{benbin}), who showed it to be a  B Ia$^+$/WNVL+OB binary. In addition to reflex RV motion in the Paschen series,
 significant  variability was present in the emission  components of profiles  obtained at comparable orbital
 phases, behaviour replicated  in the  H$\alpha$ profile (Fig. 9). Upon consideration of the full dataset 
no secular evolution was apparent in any spectral feature.

Likewise, comparison of optical  data from 2001-5 showed no long term changes for the WN9 star W44. 
However high resultion spectra revealed dramatic LPV in  H$\alpha$  and He\,{\sc i} ${\lambda}{\lambda}$ 6678, 
 7065 (Fig. 11) which, to the best of our knowledge,  are more pronounced
than the  variations observed for any other WR star outside of an LBV excursion. 
Given their absence in the spectra of our cool B 
hypergiants, the continued presence of strong He\,{\sc i}${\lambda},\lambda$ 6678, 7065  emission throughout this period 
argues against attributing the changes to a transition to a 
cool LBV phase.  On the basis of its X-ray properties
 W44 is a strong CWB candidate (Clark et al. \cite{c08}) 
and despite the above concerns  - and lack of photometric confirmation - as with W13 
we consider  binary  induced wind structure to be  the most likely explanantion for the spectral behaviour,
 but further data are 
required to confirm this hypothesis.

Finally we turn to the WC stars W239 and W241 (Fig. 12). Ritchie et al. (\cite{benbin}) showed that W239 demonstrated 
reflex binary motion between 2008-9 but,  while intrinsically variable,  the expanded dataset revealed no
evidence for secular evolution between 2001-9. Spectra of the 
dusty WC9 star W241 in 2005 May and July  demonstrated notable changes in the strengths  of  the He\,{\sc i}+{\sc ii} and 
C\,{\sc ii}+{\sc iii} lines and following the  arguements for W44 we suspect these too are due 
to binarity. Of interest are a number of sharp emission features superimposed on  the broad
C\,{\sc ii} emission line at $\sim$6580{\AA}. We suspect these reflect emission in H$\alpha$  and 
C\,{\sc ii}${\lambda}{\lambda}$ 6578, 6582, with a further, tentative identification of emission in the  
[N\,{\sc ii}]$\lambda$ 6548 line also possible
(the stronger [N\,{\sc ii}]$\lambda$6584 line is potentially  blended with C\,{\sc ii}${\lambda}$ 6582).  
The origin of such lines is currently unclear. We would not expect
narrow H$\alpha$ or [N\,{\sc ii}] emission to be associated with a WC star,  
suggesting they must originate in the hitherto unseen
companion. However, given the current data we are unable to determine whether the emission lines are nebular in 
origin or instead whether they arise in a stellar wind; if the latter is correct they would imply a rather early
 spectral type  for the secondary.

\begin{figure}
\begin{center}
\resizebox{\hsize}{!}{\includegraphics{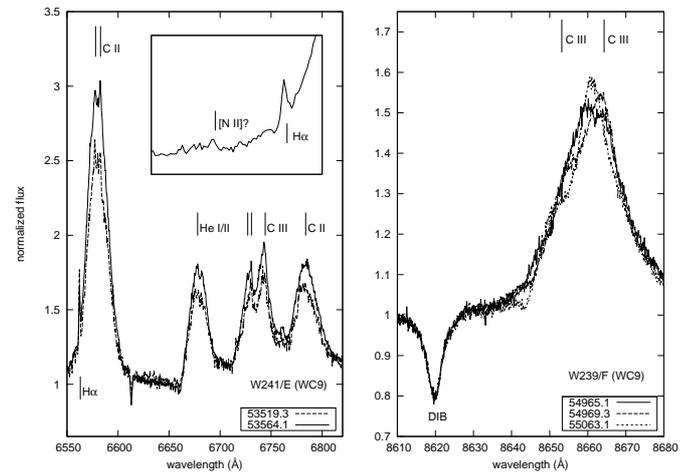}}
\caption{Variability in the spectra of the  WC9d stars W239 and W241. Note the sharp emission features
superimposed on the broad C\,{\sc ii} emission line of W241, with a possible identification  of 
 the [N\,{\sc ii}]$\lambda$  6548 line in the blue wing (indicated in the inset).}
\end{center}
\end{figure}

\section{The Yellow Hypergiants}

As with the hotter B hypergiants, field YHGs are also highly variable over multiple timescales. Most, if not all, demonstrate
 short period  pulsations with periods of the order of 10$^2$ days (e.g. de Jager \cite{dj98} and refs. therein) visible in both lightcurve and RV variations in photospheric lines; similar rapid 
variability is also present in wind contaminated emission lines such as H$\alpha$ (e.g. Lobel et al. \cite{lobel}).
 Additionally, long term secular temperature variations  have been
inferred from both spectroscopic and photometric  observations for a 
number  of the most luminous YHGs. Oudmaijer  et al. (\cite{rene96}) found that \object{IRC +10 420} increased in 
temperature by $>1000$~K over the past twenty years, with  similar behaviour inferred for
 \object{HD 179821} (Patel et al. \cite{patel}). 
In contrast \object{HR 8752} has undergone two red loops in the HR diagram 
over the past thirty years, each of approximately 10 years duration (de Jager \& Nieuwenhuijzen \cite{dj97}). 
Similar excursions to cool temperatures have been observed for \object{$\rho$ Cas} (Lobel et al. \cite{lobel})
although the duration of these events is significantly shorter ($\leq$2 yrs). 

Two or more observations of the H$\alpha$ profile are available for 5 of the 6  YHGs within Wd1. Of these 
we infer H$\alpha$ to be  variable in W265 (reported to be present by West87 but absent in our data), observe 
significant evolution in the profile of W16a (Fig. 13; apparently absent in  West87) and find no variability for W4, 8a and
 12a\footnote{Only a 
single epoch of  high resolution and S/N data is  available  for W12a - as such we would not expect to detect 
the subtle H$\alpha$ LPV observed for e.g. \object{$\rho$ Cas} (Lobel et al. \cite{lobel})}. While only 3 epochs of
spectroscopy are available for W16a, the profile appears to show a number of unusual features. The first, a 
narrow blueshifted emission component superimposed on a presumably photospheric absorption profile
 is also present in the YHGs HD 179821 (Patel et al. {\cite{patel}) and $\rho$ Cas 
(Lobel et al. \cite{lobel}) where, like here, it is variable. A broad plinth of emission 
extends to some $\pm$1000~kms$^{-1}$ and  appears constant in all three spectra, while a third emission 
component  extending from $\sim$-700kms$^{-1}$ - +300kms$^{-1}$ seems to be present in the 2002 spectrum  but 
absent in 2005; to the best of our knowledge this combination of features is not present in any other YHG. 

Pronounced emission wings are present in the mid-late B hypergiants  W7 and 42a, as well as  lower luminosity 
late B/early A field supergiants (Kaufer et al. \cite{kaufer}). Typically, such features are thought to  arise from
electron  scattering  of line photons but no strong emission is present in W16a, nor in several of the stars 
studied by Kaufer et al (\cite{kaufer}). Moreover the high velocity emission plinth  is asymmetric in W16a and is  more 
pronounced  in the red, although this might be due to blended emission in  the coincident 
[N\,{\sc ii}]$\lambda$6583 and/or C\,{\sc ii}${\lambda}{\lambda}$ 6578, 6582 lines (noting that the later appear unlikely 
given they are only seen in the extreme early B supergiants). This feature  is also 
reminiscent of the He\,{\sc i} $\lambda{\lambda}$ 5876, 6678 lines in W9 (Clark et al. \cite{c08}), which Ritchie
 et al. (\cite{ben10}) tentatively attribute to the presence of a wind collision zone. However, 
there is no evidence that W16a is a CWB; therefore  we are  currently unable to offer a convincing 
physical explanation for either the line morphology or variability of H$\alpha$ in W16a.

Preliminary observations of the neutral metallic lines of  W265 by Ritchie et al. (\cite{benbin}) suggested that it was
 pulsating; the inclusion of a  second season of observations  allows us to observe the turnaround in the  
pulsational cycle,  implying a quasi-period of order $\sim$10$^2$~days (Figs. 14-15).
Unlike the late B/early A supergiants studied by Kaufer et al. (\cite{kaufer97}), we find 
  a depth dependence for  the velocity fields of the photospheric  lines, with the higher excitation lines - which 
form deeper in the atmosphere - showing less variability than the lower excitation transitions, although all 
lines vary in phase. Comparison of the spectra taken at the extremes of this cycle 
to the classification standards of Munari \& Tomasella (\cite{MT} Fig. 16) show it to vary 
between spectral types F0 (7200K) and F5 Ia$^+$ (6570K), spending most time in the cool phase.

The magnitude of this change is directly comparable to that observed for 
$\rho$ Cas in observations between 1993-5 ($\Delta$T$\sim$750K; Lobel et al. \cite{lobel98}), although the 2000-1 outburst of $\rho$ Cas, which resulted  in a greatly enhanced mass 
loss rate, was accompanied by a more extreme evolution of  spectral type than observed here (early M supergiant; Lobel  et al. \cite{lobel}).
Moreover, {\em (i)} the timescale of quasi-pulsations (mean value $\sim$300 days; de Jager et al. \cite{djrc} and refs. therein), {\em (ii)}
the average $\Delta$RV$\sim$18kms$^{-1}$ for low excitation
(1-5eV) photospheric lines  (Lobel et al. \cite{lobel94}), {\em (iii)} the development of excess absorption in the blueshifted wings of both the Ca\,{\sc ii} and 
Fe\,{\sc i} lines during the hot phase (due to enhanced mass loss leading to an increased optical depth in an expanding atmosphere; 
Lobel et al. \cite{lobel98}, Ritchie et al. \cite{benbin}) 
 and {\em (iv)} the significant velocity stratification in
photospheric lines originating at different atmospheric depths (Lobel et al. \cite{lobel})
in $\rho$ Cas are all directly comparable to the same phenomena in W265. 
Thus both stars appear to demonstrate similar pulsationally driven mass loss, presumably via the same physical mechanism.

Unfortunately,  the spectral datasets available for W4, 8a, 12a and 16a  are of insufficient resolution and S/N to readily 
identify futher pulsators via RV shifts in the photospheric lines, although variations in the strength of 
neutral metallic lines are present in W4, and hence by analogy to W243 and 265, provide  evidence for pulsations. 
Given the relatively poor temporal sampling of these 
observations and the finding that W265 appears to spend most of its time in the cool phase, we suspect that the remaining stars may also turn out to be variable.
 Nevertheless these data  do permit accurate spectral classification\footnote{Interestingly, some evidence exists for the 
presence of abundance anomalies in these data. N\,{\sc i} absorption features are significantly enhanced in W12a with 
respect to W265;  given little temperature dependence is expected  for these lines between F1-5 Ia$^+$ we suspect this 
is the result of chemical processing. Likewise photospheric C\,{\sc i} absorption lines are present in all these stars 
but are absent in the cool-phase LBV W243 (Ritchie et al. \cite{benLBV}), suggesting C depletion in the latter object. 
 A full analysis of these data will be presented in a future paper.}; we found W4 (=F3 Ia$^+$), W8a (=F8 Ia$^+$), W12a
(=F1 Ia$^+$) and W16a (=A5 Ia$^+$) to be systematically later by 2-3 subtypes  than found by C05 utilising lower 
resolution data. Given classification of A-G supergiants is primarily  accomplished via the relative strengths of 
weak metallic lines we conclude that this discrepancy results from the different 
quality  spectra available rather than being of   astrophysical origin. A similar conclusion may also be drawn for the 
systematically  later classifications of  W4, 8a, 32 and 265  by West87 (Table 2). The classifications for W12a appear noticeably 
discrepant; we suspect this is due to an erroneously early spectral type in West87; the strength of the 
Pa18+O\,{\sc i}, Pa16+Ca\,{\sc ii} and  Pa15+Ca\,{\sc ii} blends are all directly comparable to W265 in his data, as we find 
in ours. 

Consequently, we find  no spectroscopic evidence for long term secular evolution of
 the  temperature of  any of the YHGs over the quarter of a century between 1981-2005.
Such a conclusion is bolstered
by the  photometry available; the $\sim$magnitude visual brightening found for IRC +10 420 between 1930-70 
(Humphreys et al. \cite{hump92}) would have 
been detectable if similar had occured for any YHG between 1966-2006 (Sect. 2). However, given the poor temporal 
sampling we may not exclude shorter events such as the red loops and outbursts which  \object{HR 8752} and $\rho$ Cas have been observed to 
undergo (de Jager \cite{dj98}, Lobel et al. \cite{lobel}), although long lived excursions to cool temperatures, such as seen for \object{M33 Var A} -
 for which a transition from an F to M supergiant,
 accompanied by a 3~magnitude decrease in the V band, persisted for  $\sim$45years (Humphreys et al. \cite{m33}) -
would have been detected. 

\begin{figure}
\begin{center}
\resizebox{\hsize}{!}{\includegraphics{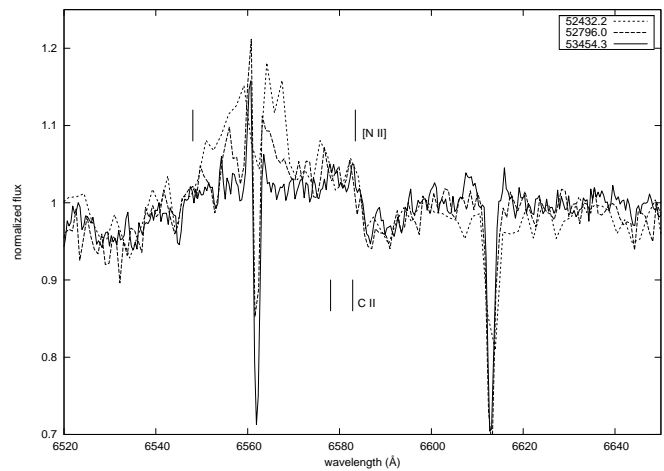}}
\caption{LPV in H$\alpha$ in W16a (A5 Ia$^+$) between 2002-5. Note the match between the strengths 
of the DIB in the 2003 \& 2005 spectra implies that the change in strength of the narrow emission and absorption
components is real. The location of C\,{\sc ii}${\lambda}{\lambda}$ 6578,82  and  [N\,{\sc ii}]${\lambda}$ 6583 are
 also indicated.}
\end{center}
\end{figure}

\begin{figure}
\begin{center}
\resizebox{\hsize}{!}{\includegraphics{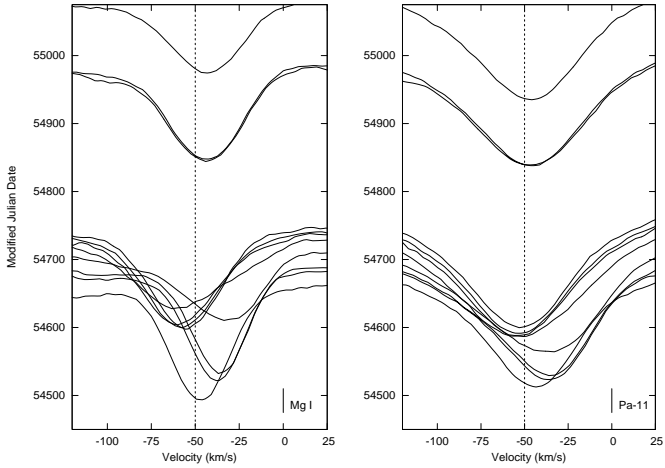}}
\caption{Pulsations in the photospheric Mg\,{\sc i} and Pa11 lines in  W265}
\end{center}
\end{figure}

\begin{figure}
\begin{center}
\resizebox{\hsize}{!}{\includegraphics{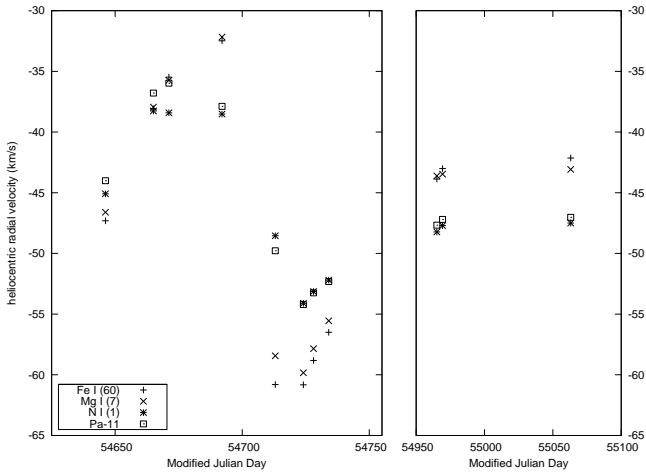}}
\caption{Plot of the evolution of the RV of selected photospheric line in W265; included are 
the upper photospheric lines of Fe\,{\sc  i} (multiplet 60, $\chi_{low} \sim$2.2~eV),
Mg\,{\sc  i} (multiplet 7, $\chi_{low} \sim$4.3~eV),
N\,{\sc  i} (multiplet 1, $\chi_{low} \sim$10.2~eV) and the Pa11 line 
($\chi_{low} \sim$12.04~eV). }
\end{center}
\end{figure}

\begin{figure}
\begin{center}
\resizebox{\hsize}{!}{\includegraphics{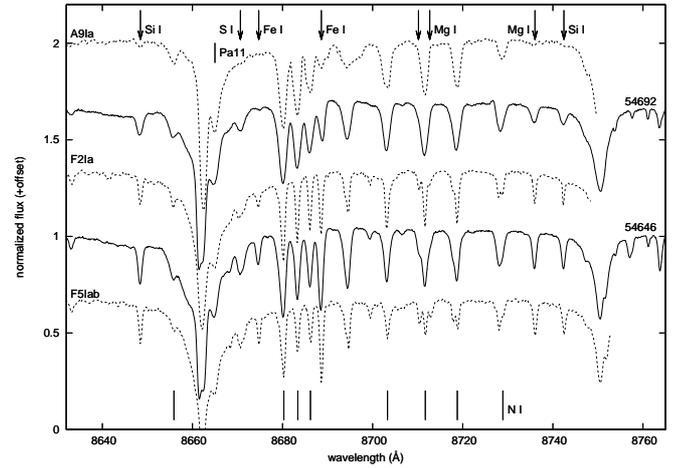}}
\caption{Comparison of spectra of W265 taken in the hot and cool phase to the classification spectra of Munari \& Tomasella (\cite{MT}).}
\end{center}
\end{figure}

\section{The Red Supergiants}

Finally we turn to the RSGs W20, 26, 75 \& 237, supplementing our sparse optical dataset (Table 5) with 
the near-IR classifications of Mengel \& Tacconi-Garman (\cite{mengel}). 
West87 report H$\alpha$ emission in W20 \& 26 and its absence in W237, with no data for W75. We fail to detect H$\alpha$ 
in W20, 
but weak emission is present in W237, although at a level that could have been missed in West87 (Fig. 17); by 
analogy with W26, we consider it likely to be a blend of H$\alpha$+[N\,{\sc ii}].
Variable line emission in the Balmer series (and low excitation species)  has previously been observed in the RSG+OB 
binary VV Cephei stars (Cowley \cite{cowley}), where it is thought to arise in the region of the RSG wind ionised by 
the hot companion. 

In contrast W26 continues to demonstrate strong H$\alpha$ emission in our data (Fig. 18),
 as well as forbidden line emission from [N\,{\sc ii}] and [S\,{\sc iii}]; given the strength of the former with respect 
to H$\alpha$   we suspect this emission arises  in nitrogen enriched ejecta rather than a stellar wind. 
 No changes are apparent in the 
strength of the H$\alpha$+[N\,{\sc ii}] emission in W26 between 2001-6, although the missmatched resolutions mean that  
we are insensitive to subtle LPV. To the best of our knowledge only 
one other  RSG - WOH G64 -  is known to show such a pronounced nebular emission  line spectrum (Levesque et al. \cite{l09}; see Appendix A for more details).

No line diagnostics  for spectral type are 
present within the wavelength regions sampled by our spectra and so we are forced to rely on the strength of the TiO bandheads
 at  $\sim$7050, $\sim$8250, $\sim$8440 \& $\sim$8860{\AA}, which increase in strength with decreasing temperature
 (e.g. Levesque et al. \cite{lev05},  Zhu et al. \cite{zhu}, Ramsey et al. \cite{ramsey}). Unfortunately, the low resolution and S/N  complicate an absolute 
classification for the spectra from 2001, but we note that the bandhead appears significantly weaker in W26 than W20 and
 W237, suggesting an earlier spectral type  for the former, while the strengths of these features in 
 W20 \& 237 clearly differ  (Figs. 17 \& 18).

Higher resolution spectra of the $\sim$8860{\AA} bandhead  for W20, W26 (both 2002 June) and  W237 (2004 June; identical to 
those from 2005 March-July) are presented in Fig. 19. These indicate that W26 is  of unambiguously earlier spectral type 
than  W20 and W237, which in these data - unlike in the  2001 spectra - are indistinguishable from one another. We therefore may conclude 
that the temperature of one or both of these stars  has varied between the two observations. 
Based on the callibration of the $\sim$8860{\AA} bandhead by Ramsey et al. (\cite{ramsey}) we may assign 
spectral types of   M2 Ia  for W26 and    M5 Ia for W20 and 237,  with an uncertainty of $\pm$1 subtype in absolute rather than relative classification. 
An additional  spectrum of W26   from 2006 February (Fig. 18) suggests a significantly later spectral type at this time in comparison to previous spectra
(no earlier than M4 Ia from the strength of the $\sim$8440{\AA} bandhead; Zhu et al. \cite{zhu}), while our sole spectrum of W75 from the same run appears to be much earlier
 than any other RSG (no later than M0 Ia; Fig. 17).

Given the different near-IR diagnostics employed by Mengel \& Tacconi-Garman (\cite{mengel})
a straightforward comparison of their classifications  to those derived from these data is  difficult. However, they find W20 and W26 to be of comparable spectral type 
(M5 Ia) and later than W237 (M3 Ia), with  their classification  of W26 being compatible with ours from the previous month. 

We therefore conclude that W26 appears to  unambiguously demonstrate  changes in spectral type (M2-5 Ia), 
with pronounced variations in the TiO bandhead diagnostics between 2001-6. W237 also appears to  be spectroscopically variable 
based its classification  relative to W20 and W26, with a comparison of the photometry presented by C05 and 
Bonanos (\cite{bonanos}) supporting this assertion (Sect.2; we cannot exclude the possibility  that W20 
is also variable). Finally,  we simply note the discrepancy between optical ($<$M0 Ia) and near-IR  (M4 Ia) 
classifications of W75 between February-March 2007.

\begin{figure}
\begin{center}
\resizebox{\hsize}{!}{\includegraphics{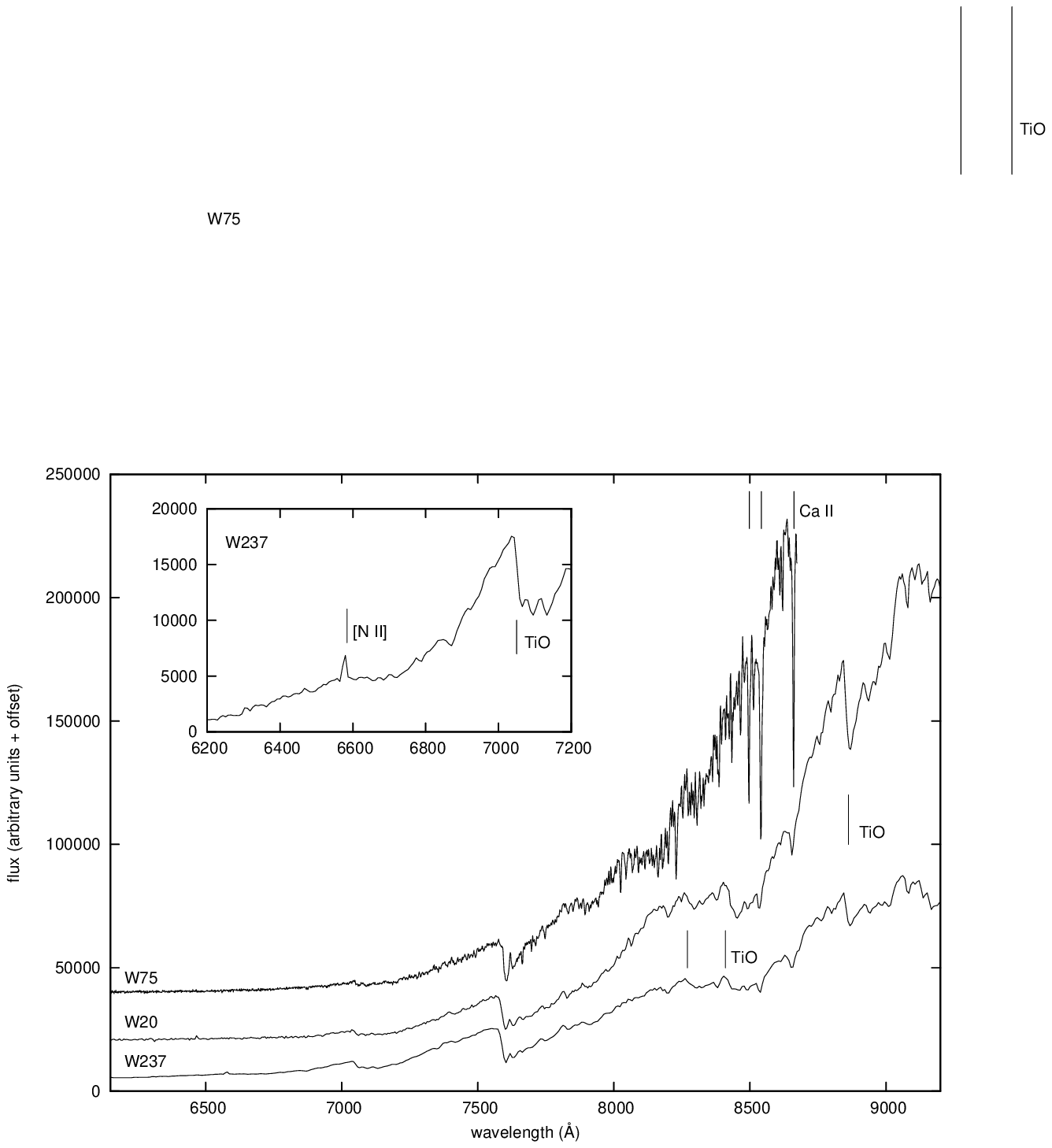}}
\caption{Low resolution spectra of the RSGs W20 \& 237  from 2001 June and a medium resolution spectrum of W75 from 2006 February, with 
an inset centred on the H$\alpha$+[N\,{\sc ii}] blend in W237. Note that no emission is present in this feature 
in W20, while the S/N is insufficient to comment for W75.}
\end{center}
\end{figure}

\begin{figure}
\begin{center}
\resizebox{\hsize}{!}{\includegraphics{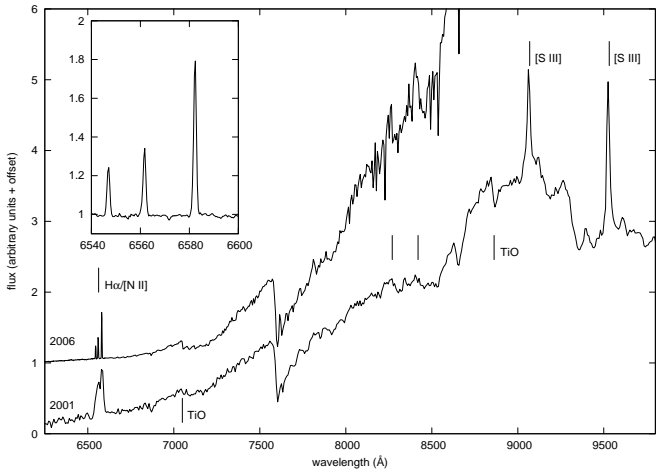}}
\caption{Spectra  of  W26 from 2001 June and 2006 February respectively; note that the resolution of the 2006 
spectrum has been artificially degraded to better match that of the 2001 data. 
The  TiO bandheads are significantly stronger in the latter spectrum, indicative of a later spectral type.
The inset shows the region of the spectrum from 2006 centred on H$\alpha$ at the original resolution.}  
\end{center}
\end{figure}

\begin{figure}
\begin{center}
\resizebox{\hsize}{!}{\includegraphics{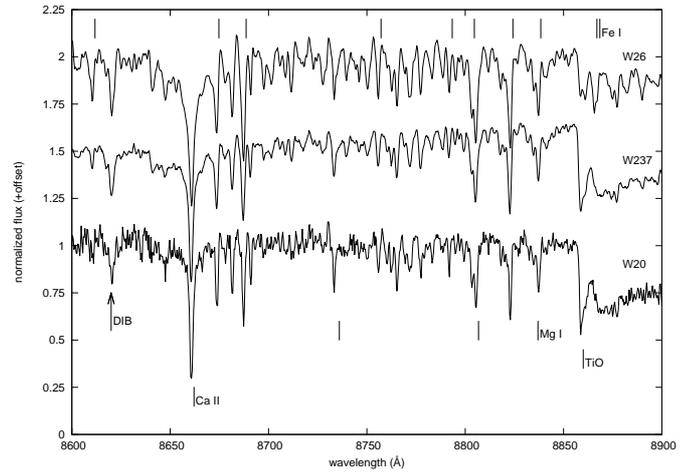}}
\caption{High resolution I band spectra of the RSGs W20, 26 (2002 June) and 237 (2004 June) covering the diagnostic  TiO 
 bandhead at 8860{\AA}.}
\end{center}
\end{figure}

Both photometric  and spectroscopic  variability may be present in high luminosity RSGs, although the latter is thought 
to be particularly rare (e.g. Levesque \cite{review}). The extreme RSG \object{VX Sgr} demonstrates `semi-regular variations' 
between M5.5-9.8~Ia, accompanied by  $\Delta$V$>$6mag  ($\Delta$I$\sim$1.2mag)
over  a $\sim$732~day quasi-period (Lockwood \& Wing \cite{lockwood}), while
 \object{S Per}  transited from  M4-7 Ia between 1971-4 (Fawley et al. \cite{fawley}) and has  also undergone
 pronounced photometric variability ($\sim$6mag) over the past century (Chipps et al. \cite{chipps}). 
Significant evolution in this behaviour is also seen; between 1889-1919 \object{VY CMa} underwent irregular
 variations with $\Delta$V$<1$mag. and a mean quasi-period of 600~d,  after which the excursions increased 
in both magnitude ($\Delta$V$\sim$3mag) and duration ($\sim$1100~d; Robinson \cite{robinson}).

Given the inevitable constrains imposed by our limited datasets, we conclude  that the variability of W26 and W237 is 
entirely consistent with the behaviour of these  extreme RSGs, implying that they too are unstable. Given the 
quasi-periodic nature of the lightcurves of VX Sgr, SPer and VY CMa, Lockwood \& Wing (\cite{lockwood})
liken  this behaviour to a high luminosity analogue of the pulsations that are present in Mira variables.
 The presence of substantial circumstellar ejecta 
around three of the four RSGs within Wd1 (Dougherty et al. \cite{d09}) which, in the case of W26, appears to be nitrogen 
enriched, supports the assertion of Massey et al. (\cite{massey}) that this instability may be associated with significant
 mass loss. Finally, as with both the BHGs and YHGs, the long term photometry (Table 2) precludes any long term excursions 
analagous to that undergone by  M33 Var A.

\section{Discussion and concluding remarks}

Despite the disparate nature of the photospheric and spectroscopic dataset accumulated for Wd1 over the past 
50 years, it presents a valuable resource to investigate the properties of the massive  stellar population 
and in particular to characterise the prevalence and nature of stellar variability. This is of particular relevance 
given the  unique census of evolved stars within Wd1, comprising examples of all post-MS 
stellar types thought to originate from stars with  M$_{initial} \sim$30-40M$_{\odot}$. 
Given the limited sampling of the combined dataset it was somewhat surprising to find that stars in {\em all} 
evolutionary states were variable, with this behaviour being 
attributable to both evolving wind structure(s) and stellar pulsations. Indeed these phenomena appear to be commonplace, with 26 of the 
OB super-/hypergiants and all but one of both the cool hypergiants and the Wolf Rayets included in this 
study being either photometrically and/or spectroscopically variable.

Wind variability was most evident in the H$\alpha$ line of the OB super-/hypergiants, with the line transiting from absorption
to emission in W70,  and pronounced P Cygni profiles developing in several other stars. Similar behaviour has been observed in
field stars although, as with this population,  the physical mechanism(s) driving the wind inhomogineities are poorly understood. 
In this respect the  detection of both photospheric pulsations and wind variability in the mid B supergiants 
W70 \& 71 and hypergiants W7 \& 42a is of considerable  interest and all four  stars would benefit from  further observations with a  higher cadence to
 investigate  potential causal links between the two phenomena (c.f. Markova et al. \cite{markova}). Moreover, 
these results again emphasise the uncertainty inherent in determining  mass loss rates for OB supergiants from a 
single epoch of H$\alpha$ spectroscopic data.

Significant wind structure is also inferred for the WRs investigated.
Given the identification  of binary markers such as the presence of hot dust, hard X-ray emission and, in the case of W13 and 239, reflex RV 
variations, we attribute this to the effect of the companion  via wind/wind interaction, noting that no secular evolution is
 present in any of the stars. This is encouraging since it suggests that detailed modeling of these variation may yield the 
underlying binary properties of the systems (e.g. Stevens \& Howarth \cite{stevens}) enabling a comparison to the less 
evolved stellar populations within Wd1, hence illuminating the effects of  binary interaction driven by stellar evolution.

Wind diagnostics are more difficult to identify and study in the cool hypergiants. Nevertheless variable high velocity
absorption in the blue wing of the metallic photospheric lines is present in W265, indicative of changes in the optical depth and hence physical 
properties of the wind. By analogy to $\rho$ Cas, we attribute this to pulsationally driven mass loss. 
Variability in the H$\alpha$ emission  line profile is inferred for W12a and directly observed in W16a, although 
the morphology of this line is complex and hence the physical origin of this behaviour is consequently uncertain. 

Variations in both profile strength {\em and} the RV of the line centroid were found for 
the photospheric lines of both the hot and cool evolved stellar populations, the combination indicative of pulsations rather than binary motion
 (e.g. Ritchie et al. \cite{benbin}).
Amongst the OB supergiants within Wd1, this behavour had previously been reported for B0.5-2 Ia stars; our results extend this to 
encompass the full spectral range present (O9-B4 Ia). Utilising the new spectroscopic callibrations of Negueruela et al. (\cite{n09}) we find that  
the presence of photometric variability (Bonanos \cite{bonanos}) appears to be a strong  function of evolutionary phase, being more prevalent amongst the early-mid B Ia stars in 
comparison to the late O Iab-II stars. This is qualitatively similar to the conclusions drawn by Fullerton et al. (\cite{fullerton}) regarding the 
occurence and magnitude of photospheric pulsations in OB stars and hence it is tempting to attribute the photometric  variability to photospheric pulsations, such as those inferred for the $\alpha$ Cygni variables. Unfortunately, we currently lack the high S/N and cadence  observations that would enable pulsational mode identification.  

 Amongst the cooler super-/hypergiants,  pulsations were inferred for stars with  spectral types ranging from  mid B through  
to mid F. The best studied example - W265 - was   observed to transit
between F0-5 Ia$^+$ over the course of a single cycle ($\sim$100~days); the combination of both timescale,  magnitude 
and atmospheric depth  velocity stratification of the pulsations  being  strongly reminiscent of those observed in
$\rho$  Cas, which have been attributed to non-radial modes (Lobel et al. \cite{lobel94}, \cite{lobel}).
 Although the data are sparse, both W26 and 237 display spectroscopic variability indicative of changes in 
spectral type, with the former 
being observed to evolve from M2-5 Ia. The finding  that at least two of the four RSGs within Wd1 are unstable and by analogy
with other extreme  RSGs, possibly pulsating is of 
considerable interest given {\em (i)} their extreme  luminosity and hence  high progenitor mass\footnote{ A lower limit to 
their progenitor  masses  may be inferred from the  cluster Main Sequence  turnoff. This is an important consideration 
given that the mass/luminosity    degeneracy for RSGs as they execute a red loop   prevents such a determination for 
field stars (Meynet \& Maeder \cite{meynet}).},  {\em (ii)} their rich circumstellar envelopes,  which provide evidence for 
significant recent mass loss, possibly driven by these instabilities and  {\em (iii)}
the rarity of this phenomenon (with only six other examples known in the Galaxy and Magellanic Clouds; 
Levesque \cite{review}).

Indeed, these preliminary  findings emphasise the important role studies of young massive clusters with multiplexing
 spectrographs will play 
in understanding the physics of high mass stars. In particular the detection of  photospheric pulsations 
extending from spectral types as 
early as O9 I, through  F5 Ia$^+$ hypergiants and potentially as late as the M5 Ia supergiants confirms the finding of e.g. Burki 
 (\cite{burki})  that stars in this region of the HR diagram are unstable, but critically 
{\em for a rich co-eval population of stars at a single metallicity rather than a 
heterogeneous field population.} Given the sample size - $\sim$80  $\sim$O8III-B2 Ia stars - future analysis of the 
full 2008-9 VLT/FLAMES dataset will provide valuable observational constraints  on the  location of the high 
temperature boundary of this region of instability in a statistically robust manner.

Furthermore, observational determination of  the amplitude and period of the pulsations as a function of 
temperature, luminosity  and evolutionary phase will provide direct  contraints on the underlying  physical mechanism(s) driving this instability
 - for example  strange- and/or g-modes (Kiriakidis et al. \cite{kiri}, Saio et al. \cite{saio}) - and, in  conjunction with 
chemical abundance analysis,  a detailed picture of the evolution of the internal structure of massive stars  
as they execute a  red loop on their way to becoming Wolf Rayets.  Similarly, high candence observations will enable an 
investigation  into the role pulsations play in both the origin of wind structure in hot supergiants and 
in driving mass loss in the cool hypergiants. Indeed, the  uniquely rich population of both hot and cool 
super-/hypergiants within Wd1 holds open the possibility of investigating the  varying contribution of  both line and 
pulsational driven  mass loss as a function of stellar temperature and luminosity as stars evolve across the HR diagram.

Relating to this, with the exception of the known LBV W243 (Ritchie et al. \cite{benLBV}), we find no compelling evidence 
for other LBV  like excursions or eruptions  within Wd1, despite both the large number of transitional stars present 
and the evidence for past episodes of  enhanced mass loss amongst a subset of  both hot and cool stars 
(Dougherty et al. \cite{d09}). H$\alpha$ LPV similar to that observed in the LBV HD160529  was found  for
 several of the B super-/hypergiants,  but the poor temporal coverage meant that we could not determine whether these 
corresponded to the  long term  secular changes that characterise LBV excursions,  or the short lived variations indicative 
of an asymmetric wind. 

However, while  similar limitations also apply to the 
(non-) detection of events comparable to the outbursts of $\rho$ Cas and the red loops of HR8752 in the YHG population,
 we would have expected to detect events comparable to the  multi-decade excursions of
 e.g.  M33 Var A, or the long term secular evolution of IRC +10 420. 
Given these results,  it is of interest that a comparable  analysis for the WNLh stars - which have been proposed  to 
be quiescent LBVs - within the Galactic Centre and Quintuplet cluster similarly reveals  little evidence for significant  
long term variability (Appendix B). Apparently transitional stars co-located with both 
`bona fide' LBVs and rapidly evolving cool hypergiants in  the HR diagram (e.g. Clark et al. \cite{c05b}) 
can enjoy significant periods of quiescence. 

These  results  are surprising given that  the majority of the stars observed within Wd1, from the extreme-B hypergiants 
through to the RSGs, appear to be low-level
variables and   only borderline-stable. As such one might naively have expected them to be prey to such large 
scale instabilities, which presumably  contribute to the lack of stars above  the HD limit. In this respect  
the YHGs and RSGs are of particular interest  since they are located at the low  temperature boundary of the HD limit and hence 
appear to represent the most massive stars that may exist in this region of the HR diagram at solar metallicities.

 Drout et al. (\cite{drout})  present detailed evolutionary predictions for the physical properties of 
such stars, which we may test against
 the population within Wd1. Our new spectral classifications  confirm the expectation that
YHGs evolving from $\sim$40M$_{\odot}$ stars should have relatively high temperatures ($>6000$~K), while the lack of 
secular variability for any of these stars over a period of $\sim$50yrs is consistent with the expectation of a 
relatively long lifetime for this phase ($\sim$10$^4$-10$^5$~yr).
However at these luminosities YHGs  are {\em not} expected to evolve to cooler temperatures, but  exactly 
this  evolution is indicated by the presence of 4 RSGs within  Wd1. 
An extention of the redwards loop to encompass a RSG phase would result in a corresponding reduction in the time spent 
as a YHG. While rapid  evolution through the  YHG  phase might be consistent with the  real time  
increase in stellar temperature observed for  IRC +10 420 and HD 179821, the continuing absence
 of this behaviour from the cool hypergiant  population of Wd1 will  place increasingly  
stringent constraints on such a scenario.

Therefore a  synthesis of {\em (i)} the current physical properties of the evolved stars within Wd1 - such as 
chemical composition, luminosity, temperature and, where observable, surface gravity, {\em (ii)} the relative populations of 
the different evolutionary classes and {\em (iii)} the  characterisation of their (pulsational) instabilities and secular variability
has great potential for constraining the physics of massive post-MS evolution.

Finally, regarding the ensemble properties of Wd1,  constraining the occurence and amplitude of the pulsationally 
(and binary) driven  RV changes will be crucial in allowing the accurate determination of dynamical masses for unresolved
 extragalactic clusters (e.g. Gieles et al. \cite{gieles}), where the velocity dispersion attributable to such processes may 
match or even dominate  virial motions. Likewise determining the properties of 
photometric outbursts driven by pulsational instabilities in  both hot and cool hypergiants will be
important in determining the validity of photometric cluster  mass estimates,   
particularly for the lower range of masses where the presence of such an event may dominate the integrated light.

\begin{acknowledgements}
JSC acknowledges the support of an RCUK fellowship. This research is partially supported by the Spanish Ministerio de
Ciencia e Innovaci\'on under grants AYA2008-06166-C03-03 and
Consolider-GTC CSD2006-70.
 We thank Otmar Stahl for kindly making his spectra of HD160529 available to us, and Ornette Coleman for inspiration during the 
writing of this paper. 
\end{acknowledgements}

\appendix
\section{The emission line spectrum of  the RSG W26}

Of the four RSGs within Wd1, only W26 demonstrates a pronounced nebular emission line spectrum, with emission in 
H$\alpha$, [N\,{\sc ii}]  ${\lambda}{\lambda}$6548,6582 and [S\,{\sc iii}] ${\lambda}{\lambda}$9069,9532. The
[S\,{\sc iii}] lines are common in high temperature H\,{\sc ii}  regions as well as tracing shocked
 material. No oxygen emission  - such as [O\,{\sc i}] ${\lambda}$ 6300  - is seen, while 
[S\,{\sc iii}] ${\lambda}$ 6312 is particularly 
temperature dependent (e.g.  Osterbrock \& Ferland \cite{ost}) and so its  absence while [S\,{\sc iii}]  
${\lambda}{\lambda}$9069,9532 emission is present is explicable if the region is not highly excited.
The only other RSG to demonstrate a nebular emission line spectrum is the extreme system WOH G64 which, like W26,
is also associated with an extensive dusty circumstellar envelope (Levesque et al. \cite{l09}).
 Unlike W26, WOH G64  lacks [S\,{\sc iii}] emission although [O\,{\sc i}] ${\lambda}$ 6300 
and [S\,{\sc ii}] ${\lambda}$ 6713 are present, as  is  the high excitation line 
[O\,{\sc iii}] ${\lambda}$ 5007, which our spectra do not reach (Levesque et al. \cite{l09}).

Given the lack of many (blue) nebular diagnostic lines  we are unable to present  quantitative conclusions
as to the chemical abundances or excitation mechanism for the emission spectrum. However, by analogy to both WOH G64 and the 
nebula around the BSG Sher25, we suspect the strength of the [N\,{\sc ii}] emission implies N enrichment in the circumstellar
material, presumably as the result of CNO processing (Levesque et al. \cite{l09}). Likewise, while we suspect the 
[S\,{\sc iii}]  emission is shock excited, we cannot exclude  the possibility of excitation by either 
a hot companion or the  diffuse UV radiation field of the cluster; indeed the location of 
W26 within the 'core' of Wd1 suggests that the latter may well play a significant role (c.f. Dougherty et al. \cite{d09}). Nevertheless, 
the lack of comparable nebular emission in W20 and 237 is notable, particularly given that these 2 RSGs also possess
extended circumstellar envelopes comprising both dust and ionised gas; albeit significant less massive and dense than found for 
W26 (Dougherty et al. \cite{d09}).

\section{Long term stellar variability in the Galactic Centre and 
Quintuplet clusters}
 
Comparable variability analyses using published data may also
be undertaken for both the Galactic Centre (GC - 6~Myr; Paumard et al. \cite{paumard}). 
and Quintuplet clusters ($\sim$4Myr; Figer et al. \cite{figer}).
Multiple spectroscopic observations  spanning a decade have been made of the emission line star population 
 within the GC cluster (e.g. Najarro et al. \cite{najarro},  Paumard et al. \cite{paumard01}, Martins et al. 
\cite{martins}) which are complemented by the long term ($\sim$decade)  
photometric datasets of Rafelski et al. (\cite{rafelski}) and  Trippe et al. (\cite{trippe}).
In the absence of   cool hypergiants  and comparable multiple spectroscopic 
datasets   for the OB supergiant population
 we restrict the discussion to the narrow line He\,{\sc i} stars,  which recent analysis by Martins et al. 
(\cite{martins}) confirm  to be Ofpe/WNL stars. Despite the  implication of such stars in the LBV
 phenomena only one star - IRS34W -  was observed to undergo photometic variations of a magnitude comparable to canonical 
LBV excursions (Trippe et al. \cite{trippe}), with a second, IRS16NW, being a low level variable 
($\Delta$K$\sim$0.2mag). However, with the exception of reflex radial velocity motion in the eclipsing binary IRS16SW,
 no spectroscopic variability was found for {\em any} of these stars, including IRS34W (Trippe et al. \cite{trippe};
Martins, priv. comm. 2009), noting that the S/N and resolution of the spectra would clearly have identified any 
significant changes in morphology characteristic of the LBV phenomenon (e.g. Clark et al. \cite{c09}), as well as the 
LPV exhibited by e.g. W7 and W33.

Regarding the Quintuplet, both the cool  B hypergiants FMM362 and the Pistol star are spectroscopically 
 (Figer  et al. \cite{figer}; Figer, priv. comm. 2009) and photometrically variable (Glass et al. \cite{glass}), 
suggesting that they are undergoing LBV excursions. Unfortunately, only two epochs of spectroscopy are 
available for the remaining cluster population, separated by  of order a decade (Figer et al. \cite{figer}, Liermann et al. \cite{liermann}). Of these,
 the former are of  significantly lower resolution and S/N, although are still sufficient to distinguish between e.g. 
cool hypergiants such as the Pistol star, hotter Ofpe/WNL and $\sim$featureless OB supergiant spectra. However, with the 
exception of  FMM362 and the Pistol star,  none of the other stars in common between the studies appear
 to demonstrate  unambiguous spectral variability,  
despite the presence of a number of luminous WN9  and  late O/early B supergiants within the Quintuplet. 
These results are bolstered by Glass et al. (\cite{glass}) who  only found two other confirmed cluster members 
to be photometrically variable between 1994-7; the dusty WCL stars Q2 and Q3, for which this behaviour may be attributed to 
time variable dust production.

Thus, somewhat suprisingly, of the 10 WN9 and WN9/Ofpe stars for which  
multiepoch data exist, none have demonstrated  characteristic LBV spectral variability, while only a single object would have been 
photometrically classified as such.  Mindful of  the sparse nature of the Quintuplet datasets, these 
findings are  {\em suggestive} of a rather long duty cycle for such excursions if such stars are quiescent hot phase 
LBVs.

\end{document}